\documentclass[twocolumn]{aastex61} 

\usepackage{graphicx}
\usepackage{natbib}
\usepackage{multirow} 
\usepackage{textcomp}
\usepackage{lipsum}
\usepackage{ulem}
\usepackage{chngcntr}

\usepackage{color}

\newcommand{\bc}{\begin{center}}
\newcommand{\ec}{\end{center}}

\newcommand{\hii}{H {\scriptsize II}}
\newcommand{\hi}{H {\scriptsize I}}
\newcommand{\msun}{M$_{\odot}$}
\newcommand{\lsun}{L$_{\odot}$}

\newcommand{\ra}{$\alpha$(2000)}
\newcommand{\dec}{$\delta$(2000)}
\newcommand{\h}{$^{\mathrm{h}}$}
\newcommand{\m}{$^{\mathrm{m}}$}
\newcommand{\s}{$^{\mathrm{s}}$}

\newcommand{\kms}{km s$^{-1}$}
\newcommand{\kmsp}{km s$^{-1}$ pc$^{-1}$}

\newcommand{\jyb}{Jy beam$^{-1}$}

\newcommand{\pdot}{$\dot\textrm{p}_{rad}$}
\newcommand{\prad}{$\textrm{p}_{rad}$}
\newcommand{\psn}{$\textrm{p}_{sn}$}
\newcommand{\tage}{$t_{age}$}

\newcommand{\negfifteen}{-$15$}

\newcommand{\cii}{[C{\scriptsize II}]}
\newcommand{\am}{NH$_{3}$}

\newcommand{\water}{H$_2$O}
\newcommand{\rrl}{H64$\alpha$ and H63$\alpha$}
\newcommand{\rrll}{H64$\alpha$, and H63$\alpha$}

\newcommand{\hco}{H$^{13}$CO$^{+}$}

\newcommand{\htwo}{H$_2$}
\newcommand{\pa}{P$\alpha$}

\newcommand{\sick}{the Sickle H {\scriptsize II} region}
 
\newcommand{\scloud}{M0.20$-$0.033}
\newcommand{\pistol}{the Pistol nebula} 
\newcommand{\pistoln}{G0.15$-$0.05} 
\newcommand{\ncloud}{M0.17$-$0.08}
\newcommand{\qc}{Quintuplet cluster} 
\newcommand{\rar}{Radio Arc region} 
\newcommand{\Rar}{The Radio Arc region} 

\newcommand{\sickp}{the Sickle H {\scriptsize II} pillars}
\newcommand{\slbv}{G0.120$-$0.048}
\newcommand{\low}{low-velocity} 
\newcommand{\loww}{low-} 
\newcommand{\high}{high-velocity} 
\newcommand{\Loww}{Low-} 
\newcommand{\High}{High-Velocity} 
\newcommand{\mshell}{\scloud~expanding shell}
\newcommand{\gcloud}{M0.10$-$0.08}

\newcommand{\twenty}{M$-$0.13$-$0.08}

\newcommand{\sgra}{Sgr A$^*$}
\newcommand{\irbubble}{Radio Arc bubble}

\newcommand{\regs}{regions}
\newcommand{\rega}{region A}

\newcommand{\Regb}{Region B}
\newcommand{\regb}{region B}
\newcommand{\regc}{region C}
\newcommand{\Regc}{Region C}
\newcommand{\vsys}{$v_{sys}$}
\newcommand{\vexp}{$v_{exp}$}
\newcommand{\vmin}{$v_{min}$}
\newcommand{\vmax}{$v_{max}$}

\newcommand{\pv}{position-velocity}
\newcommand{\PV}{Position-Velocity}

\newcommand{\mc}{MC} 
\newcommand{\mcs}{MCs} 
\newcommand{\gc}{GC} 
\newcommand{\lbv}{luminous blue variable}

\newcommand{\xray}{X-ray}
\newcommand{\los}{line-of-sight}
\newcommand{\beam}{restoring beam}

\newcommand{\ie}{i.e.}
\newcommand{\eg}{e.g.}
\newcommand{\til}{$\sim$}
\newcommand{\andd}{\&}
\newcommand{\leftit}{\textit{left}}
\newcommand{\rightit}{\textit{right}}

\begin{document}


\title{\scloud: An Expanding Molecular Shell in the Galactic Center Radio Arc} 

\author{Natalie Butterfield} 
\affil{Department of Physics \& Astronomy, University of Iowa, 203 Van Allen Hall, Iowa City, IA 52242, USA}
\email{natalie-butterfield@uiowa.edu}

\author{Cornelia C. Lang}
\affil{Department of Physics \& Astronomy, University of Iowa, 203 Van Allen Hall, Iowa City, IA 52242, USA}

\author{Mark Morris}
\affil{Department of Physics \& Astronomy, University of California, 430 Portola Plaza, Los Angeles, CA 90095, USA} 

\author{Elisabeth A.C. Mills}
\affil{Department of Physics \& Astronomy, San Jose State University, 1 Washington Square, San Jose, CA 95192, USA}

\author{Juergen Ott}
\affil{National Radio Astronomy Observatory, 1003 Lopezville Road, Socorro, NM 87801, USA}

\begin{abstract}
We present high-frequency Karl G. Jansky Very Large Array (VLA) continuum and spectral line (\am, \rrll) observations of the Galactic Center~Radio Arc region, covering \sick, the \qc, and molecular clouds \scloud~and \gcloud. These observations show that the two velocity components of \scloud~(\til25 \andd~80 \kms), previously thought to be separate clouds along the same \los, are physically connected in \pv~space via a third southern component around 50 \kms. 
Further \pv~analysis of the surrounding region, using lower-resolution survey observations taken with the Mopra and ATCA telescopes, indicates that both molecular components in \scloud~are physically connected to the \gcloud~molecular cloud, which is suggested to be located on stream 1 in the \cite{Kru15} orbital model. 
The morphology and kinematics of the molecular gas in \scloud~indicate that the two velocity components in \scloud~constitute an expanding shell. 
Our observations suggest that the \mshell~has an expansion velocity of 40 \kms,~with a systemic velocity of 53 \kms, comparable to velocities detected in \gcloud. 
The origin of the expanding shell is located near the \qc, suggesting that the energy and momentum output from this massive stellar cluster may have contributed to the expansion. 
\end{abstract}

\keywords{Galaxy: center, ISM: kinematics and dynamics, ISM: bubbles}

\section{Introduction}

The central molecular zone of the Galactic center (\gc; a region \til200 pc in extent) is known to be an extreme Galactic environment. Molecular clouds (\mcs) in the \gc~are believed to have hotter gas temperatures \citep[50$-$300 K;][]{Mauers86, M+M13, krieger16}, higher densities \citep[$10^{4-5}$ cm$^{-3}$;][]{Zylka92}, and broader line widths \citep[\til20$-$50 \kms;][]{Bally87} than clouds in the disk \citep[$\sim$10$-$20 K, $10^{2-3}$ cm$^{-3}$, \til1$-$10 \kms;][]{Huttem93b, oka01a, Longmore13a}. The connection between these individual clouds and the larger structure of molecular gas in the \gc~is an area of active study \citep{Molinari11, Kru15}. 
The molecular gas has been modeled as several `orbital streams' by using the central velocities of the individual gas clouds and projecting their distances from \sgra.
In some cases, the complex kinematics of the molecular gas in the \gc, caused by local sources of energy injection,~makes modeling the overall distribution of the molecular gas challenging. 

The \gc~Radio Arc is located $\sim$25 pc in projection from \sgra~ and is one of the regions where modeling the orbital stream is difficult. \Rar~includes an extended \mc~\citep[\scloud;][]{SG91,serabyn94}, the Sickle \hii~region \citep[G0.18$-$0.04;][]{YZ87,lang97}, and the \qc~\citep{figer99a}. The Lyman continuum ionizing rate of \sick~(2.8$\times$10$^{49}$ photons s$^{-1}$) indicates the \hii~region is the result of ionization from the adjacent \qc~heating the edge of \scloud~\citep{SG91,lang97}. The southern extent of the \rar~includes the core of a dense \mc~(\gcloud). Single-dish ($\sim$35$-$18\arcsec~resolution) observations of \hco~(J=1$-0$) and SiO (J=1$-0$) show that \gcloud~is a denser clump of a larger diffuse cloud \citep{Tsuboi97,Handa06, Tsuboi11}.

\begin{figure}
\includegraphics[scale=0.43]{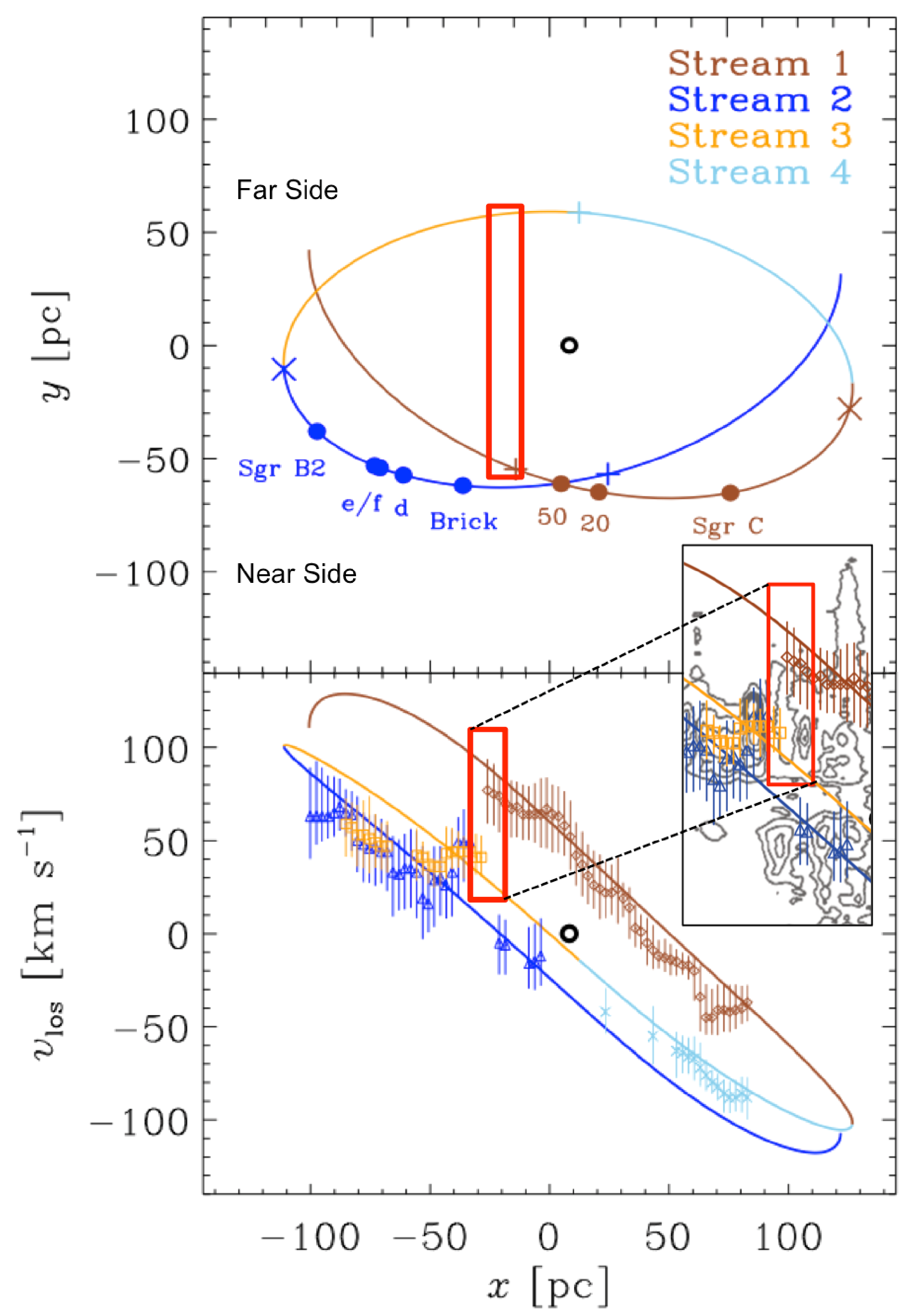}
\caption{Orbital model from Figures 6 \andd~4 in \cite{Kru15}, showing the spatial orientation ({\it top}) and velocity distribution ({\it bottom}), respectively. In their model the \gc~molecular gas has an open orbit solution that is divided into four orbital streams. The data points in the {\it bottom} panel show the observed molecular gas velocities compared to the modeled orbital streams (solid line). The red box in both panels shows the velocities ({\it bottom}) and locations ({\it top}) of the two molecular components in \scloud. 
({\it inset}): \cii~contours from \cite{langer17} (cropped from their Figure 7), overlaid on the \cite{Kru15} orbital model. }
\label{orbitalmodel}
\end{figure}

Figure \ref{orbitalmodel} (top) illustrates the orbital model proposed by \cite{Kru15} (their Figure 6). Figure \ref{orbitalmodel} (bottom) shows the molecular gas velocities (data points) compared to the \cite{Kru15} orbital model (solid lines) from their Figure 4. At the location of the \rar~there are two velocity components in the molecular gas, at \til25 \andd~80 \kms~(\scloud, red box in Figure \ref{orbitalmodel}, bottom). \cite{Kru15} attributes these two components to being on the near side (80 \kms; stream 1) and far side (25 \kms; stream 3) of the \gc~(red box in Figure \ref{orbitalmodel}, top).

We use high-resolution Karl G. Jansky Very Large Array (hereafter, VLA) observations, a facility operated by the National Radio Astronomy Observatory (NRAO),\footnote{The National Radio Astronomy Observatory is a facility of the National Science Foundation operated under cooperative agreement by Associated Universities, Inc.} to investigate the complex kinematics in the \rar~and how massive star clusters influence the \gc~ISM. In this paper we present the 24.5 GHz  continuum and spectral line results for the \rar~(Sections \ref{sconttext} \andd~\ref{spec}, respectively).\footnote{A more detailed discussion of the \gcloud~\mc~will be presented in a forthcoming paper (Butterfield et al. 2017, in prep)} All results presented in this paper are discussed in Section \ref{dis}.


\section{Observations and Data Calibration} 
\label{obs} 

\subsection{Observational Set up}
\begin{figure*}
\includegraphics[scale=0.74]{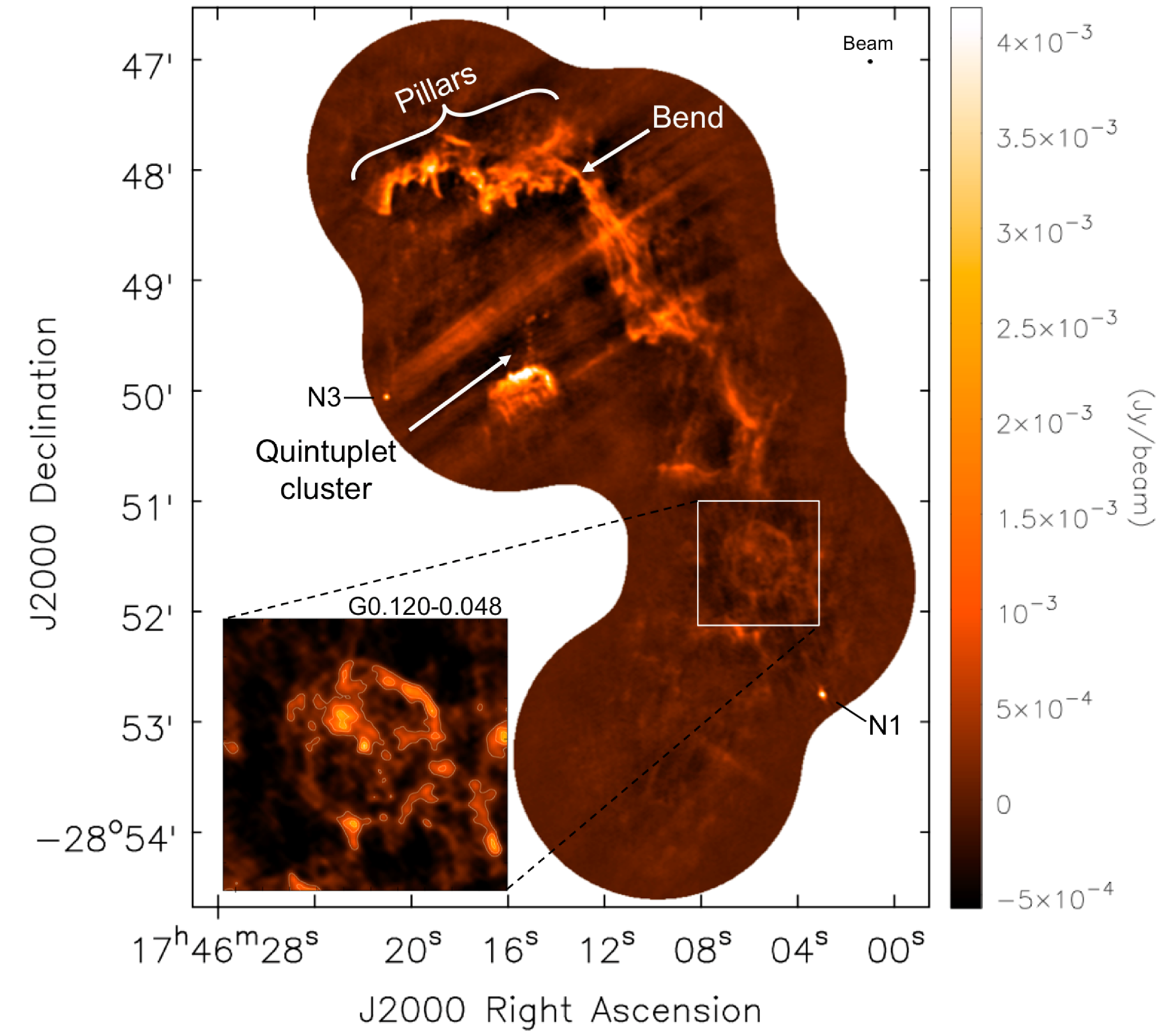}
\caption{24.5 GHz continuum image of \sick~and \gcloud~(southernmost field), using 2 GHz bandwidth. Labeled are several features discussed in Section \ref{sconttext}. ({\it inset}): The \lbv:~\slbv, at higher contrast, with grey contours showing: 4, 7, \andd~10 $\times$ 50 $\mu$\jyb~(rms noise). The \beam, shown in the top right corner, is $2\farcs18$$\times$$1\farcs86$, PA=82$\fdg$8.}
\label{sicklecont}
\end{figure*}

We used the WIDAR correlator of the VLA to simultaneously obtain wide bandwidth continuum and multiple spectral lines in the \rar. The data were taken with the K band receiver (18.0$-$26.5 GHz) on 2012 January 14. All observations used the hybrid DnC array in order to compensate for the low altitude of the \gc~at the VLA site. The radio observation subbands presented in this paper were centered at 24.054 and 25.374 GHz, each with a bandwidth of 0.84 GHz that is segmented into seven spectral windows. Two additional spectral windows were observed at X band (\til8 GHz) to determine the pointing accuracy during the observations. We had 512 channels per spectral window in each subband, with a corresponding spectral resolution of 250 kHz (\til3 \kms). This spectral configuration enabled us to accurately observe the \am~(3,3), \rrll~spectral lines. The \am~(3,3) transition (hereafter, \am), has a rest frequency of 23.87013 GHz. The \rrl~radio recombination lines have a rest frequency of 24.50990 GHz and 25.68628 GHz, respectively. The primary beam FWHM is \til$3\farcm4$ at K band, therefore six Nyquist sampled pointings were needed to survey the region with a total of $\sim$25 minutes on source per pointing.\footnote{Figures \ref{sicklecont} and \ref{all am} show our total field of view for the VLA observations presented in this paper.}

\subsection{Calibration and Imaging} 

We used the Common Astronomy Software Application (CASA)\footnote{\href{url}{http://casa.nrao.edu/}} program provided by NRAO to perform all rudimentary calibration steps (\ie, bandpass, flux, and phase corrections), as well as correcting for atmospheric opacity and antenna delay solutions. These additional corrections were needed due to the increased water vapor opacity at higher VLA frequencies. These observations used 3C286 as the absolute flux calibrator and J1733-1304 as the bandpass calibrator. Our phase calibrator, J1744-3166, was observed at 25 minute intervals. 
 
All continuum and molecular transitions were imaged using the CASA task CLEAN. Since all of the regions surveyed required multiple pointings, the CLEAN parameter `mosaic' was used to combine the observed fields. The continuum images were obtained by flagging out all spectral lines and end channels. We used `briggs' weighting with a `robust' parameter of 0.5 to balance our point-source resolution ($1\farcs86$$\times$$2\farcs18$)\footnote{Where 1\arcsec~is 0.04 pc at a distance of 8.0 kpc to the \gc~\citep{boehle16}} with the sensitivity of the image (50 $\mu$\jyb, rms noise). 

The spectral lines were first continuum subtracted using the CASA task IMCONTSUB on line-free channels. Spectral lines were imaged using their natural frequency resolution with no spectral smoothing and `natural' weighting for all transitions. The spectral images for \am, \rrll~were then spatially smoothed from their natural resolution to boost the low signal-to-noise ratios (SNRs) in the \rar,~resulting in a spatial resolution of 5\farcs0$-$5\farcs7. Due to the extended nature of the \am~emission, we used the CLEAN parameter `multiscale' to maximize the sensitivity to large-scale diffuse structures in the images (where the largest angular size is $\sim$60\arcsec). The \rrl~recombination lines were averaged in the image plane to boost our SNR, using the CASA task IMMATH. 


\section{Radio Continuum Emission} 
\label{sconttext}

Figure \ref{sicklecont} shows the 24.5 GHz continuum emission in the Radio Arc and \gcloud~regions. The large-scale curved structure (\til5\arcmin; 12 pc) in the four northern fields is \sick~\citep[G0.18$-$0.04;][]{YZ87, lang97}. The Sickle \hii~region has several distinct features (indicated in Figure \ref{sicklecont}): the `Sickle \hii~pillars,' the `Bend,' and the striations associated with the non-thermal filaments (NTFs). The radio continuum emission in \sick~is clumpy in nature, with several knots of emission that are oriented along the southern edge of the pillars region. Several of these clumps protrude from \sick~towards decreasing declination, resulting in the pillar-like appearance. These pillar-like features are \til10$-$20\arcsec~(0.4$-$0.8 pc) in length, with a width of $\sim$2$-$10\arcsec~(0.08$-$0.4 pc). Many of these pillars show a continuum brightening near one end. The $\sim$$90\degr$ Bend in \sick~appears to separate the clumpy material in the Sickle \hii~pillars from the southern striations. 

We also detect several Radio Arc NTFs at 24.5 GHz \citep{YZ87}. These NTFs are oriented perpendicular to the western portion of the \sick~and Galactic plane, as seen in Figure \ref{sicklecont}. At the apparent intersection of the NTFs and \sick, there are several bright, compact knots of emission. Located along one of these NTFs is \pistol~\citep{figer99b}, which contains the brightest diffuse emission in the entire \rar. North of \pistol~are several point-like radio sources, produced by the stellar winds of massive stars from the \qc~\citep{lang05}. 

Located near the edge of our primary beam are two bright point sources: N1 and N3. These two unresolved sources contain the brightest compact emission in the Radio Arc region. When comparing the radio emission to Paschen-$\alpha$ (hereafter, \pa) emission in \cite{Wang10}, N1 has a \pa~counterpart, indicating that it is thermal in nature. N3 does not have a \pa~counterpart, suggesting that it is non-thermal in nature. This is consistent with the high frequency (10$-$49 GHz) spectral index of N3 \citep[$\alpha = -$0.8;][]{dom16}.

\begin{figure*}[tb!]
\includegraphics[scale=0.65]{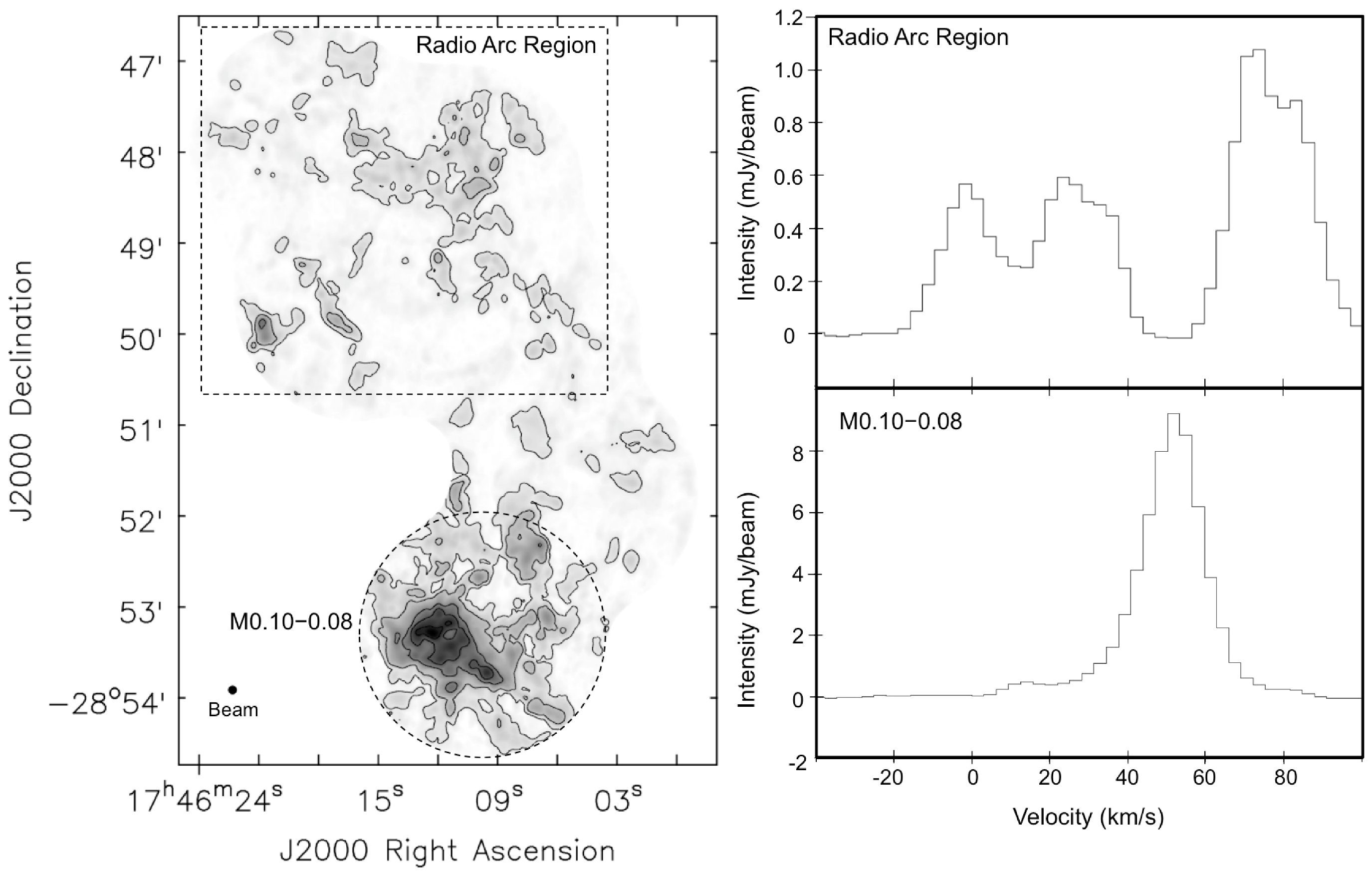}
\caption{({\it left}): maximum intensity \am~(3,3) emission ($5\farcs0 \times 5\farcs0$, PA=0$\fdg$0, resolution), integrated over a velocity range of -20$-$100 \kms, with contours showing: 5, 15, 50, 100, \andd~200 $\times$ 3 m\jyb~(rms noise). ({\it right}): spectra integrated over the outlined regions shown on the {\it left} for the \rar~({\it top}) and \gcloud~({\it bottom}). } 
\label{all am}
\end{figure*}

The \gcloud~\mc~is located in the southernmost field in Figure \ref{sicklecont} and has very low-level continuum compared to the rest of the emission in the image. There is a very faint filamentary streak near the right edge of this field. At 0.3 m\jyb, this feature is the only emission in this field that is detected above 5$\sigma$. This continuum feature is also seen in the \pa~emission from \cite{Wang10}, indicating that it is thermal in nature.

We also present the first radio detection of the recently discovered, isolated, \lbv: \slbv~\citep{MauLBV10, Lau14} and surrounding shell (Figure \ref{sicklecont}, inset). Unlike the ejected material from the Pistol star (\pistol; \pistoln), the shell surrounding \slbv~has a circular structure that is also seen in \pa~\citep{MauLBV10}. This circular structure suggests that the shell has not strongly interacted with the asymmetrically distributed ISM material that would distort the symmetric shape. The \slbv~shell is 37\arcsec~(1.4 pc) in diameter and shows a slight continuum brightening on the northern edge.


\begin{figure*}
\includegraphics[scale=0.71]{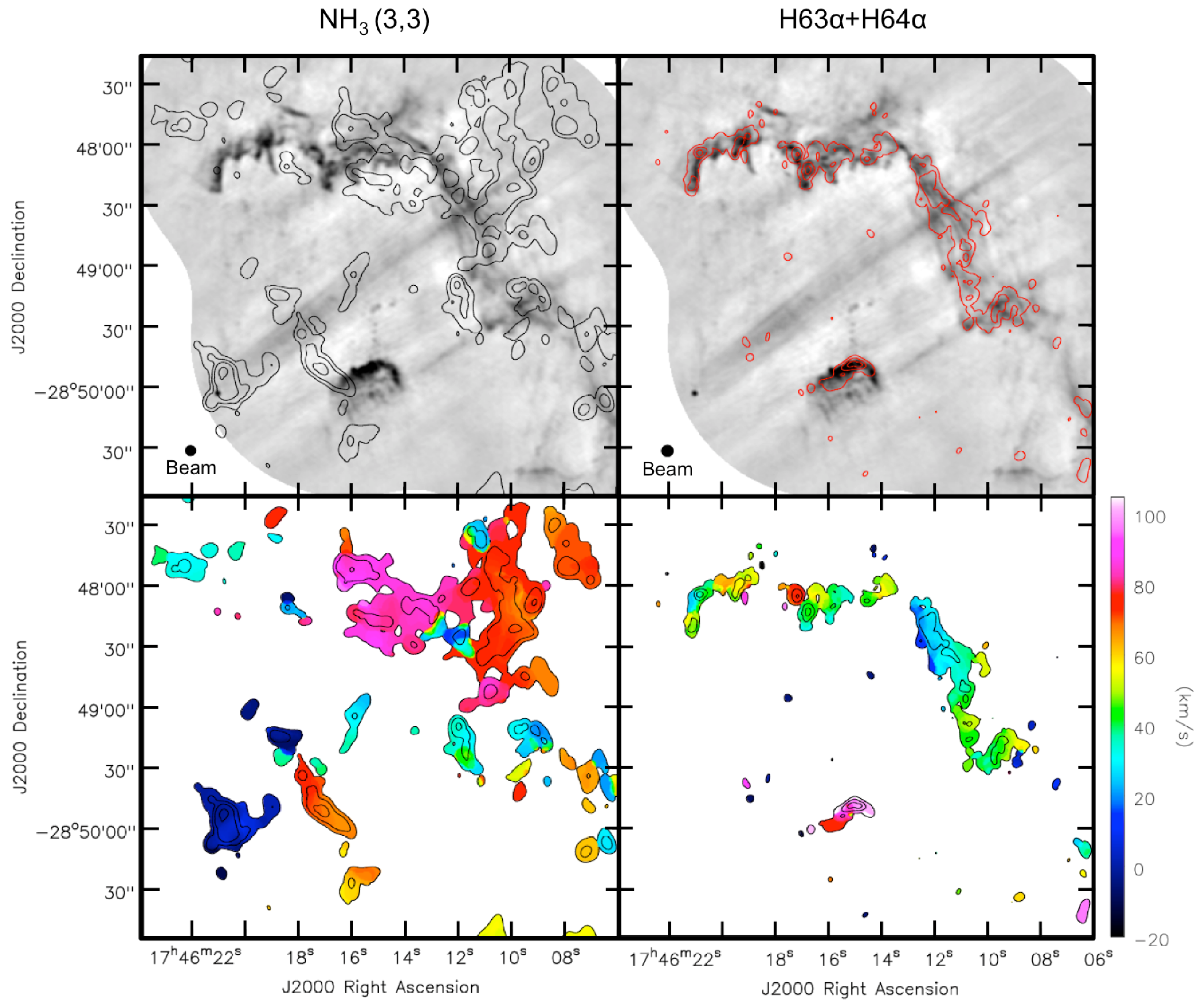}
\caption{Distribution of \am~(3,3) ($5\farcs0$$\times$$5\farcs0$, PA=0$\fdg$0, resolution, {\it left} panels) and combined H63$\alpha$+H64$\alpha$ radio recombination lines ($5\farcs68$$\times$$5\farcs61$, PA=88$\fdg$5, resolution, {\it right} panels). The {\it top} panels show maximum intensity contours in: \am~at 5, 10, 20, \andd~50 $\times$ 3 m\jyb~(rms level), and the combined H63$\alpha$+H64$\alpha$ radio recombination lines at 5, 6, 8, \andd~10 $\times$ 1.1 m\jyb~(rms level) overlaid on the 24.5 GHz radio continuum, shown in greyscale. The {\it bottom} panels show the same maximum intensity contours, now overlaid on the intensity-weighted velocity distribution for emission between -20$-$100 \kms.}
\label{ra mole}
\end{figure*}

\section{Spectral Line Emission}
\label{spec}
 
Figure \ref{all am} (left) shows widespread \am~emission throughout our surveyed regions, where the brightest emission is associated with \gcloud. The \am~emission in \gcloud~is over five times brighter than emission detected in the \rar. Figure \ref{all am} (right) shows spectral profiles integrated over the two regions indicated in the emission map. The \am~spectrum of the \rar~shows three velocity components: -15$-$15 \kms, 15$-$40 \kms, and 60$-$90 \kms~(Figure \ref{all am}, top right). The 15$-$40 \kms~and 60$-$90 \kms~molecular components are known to be associated with \scloud~\citep{SG91}. Hereafter, the molecular gas features associated with the 15$-$40 \kms~and 60$-$90 \kms~emission will be identified as the `\low'~and `\high'~molecular components, respectively. The \gcloud~spectrum (Figure \ref{all am}, bottom right) shows most of the \am~emission is contained within a single component that ranges in velocity from \til35$-$70 \kms, that peaks in intensity around 55 \kms. This velocity has an intermediate value between the \loww~and \high~molecular components in \scloud. 

Figure \ref{ra mole} presents the \am~(left) and the combined \rrl~radio recombination lines (right) in the \rar. The top panels in Figure \ref{ra mole} show the distribution of the spectral line emission with respect to the 24.5 GHz continuum features. The bottom panels in Figure \ref{ra mole} show the velocity distribution of the spectral line emission. 

\subsection{Morphology and Kinematics of Molecular Gas}
\label{ra morph}

Most of the \am~emission in the \rar~is associated with \scloud, as shown by the velocity distribution in Figure \ref{ra mole} (bottom left). The velocity distribution shows that most of the compact \am~emission in \scloud~is associated with the \high~molecular component, consistent with the integrated spectrum (Figure \ref{all am}, top right). The largest emission region in the high velocity molecular component is $\sim$2$\arcmin$~(4.7 pc) in length and spatially overlaps with the western side of \sickp~(Figure \ref{ra mole}, top left). Very little \am~emission is detected near the eastern side of \sickp. The emission associated with the \low~molecular component of \scloud~(Figure \ref{ra mole}, bottom left) is distributed in low level ($<$10$\sigma$) disjointed clumps that are $\sim$30\arcsec~or less in extent. The \am~emission associated with the~\negfifteen$-$15 \kms~molecular component is primarily located in the N3 \mc~\citep[\ncloud;][]{dom16}. The \am~emission in the N3 \mc~is the brightest emission detected in Figure \ref{ra mole}.  There are also a few, faint, isolated clumps of emission at this velocity range that are located between the N3 \mc~and \sickp.

\begin{figure*}
\includegraphics[scale=0.65]{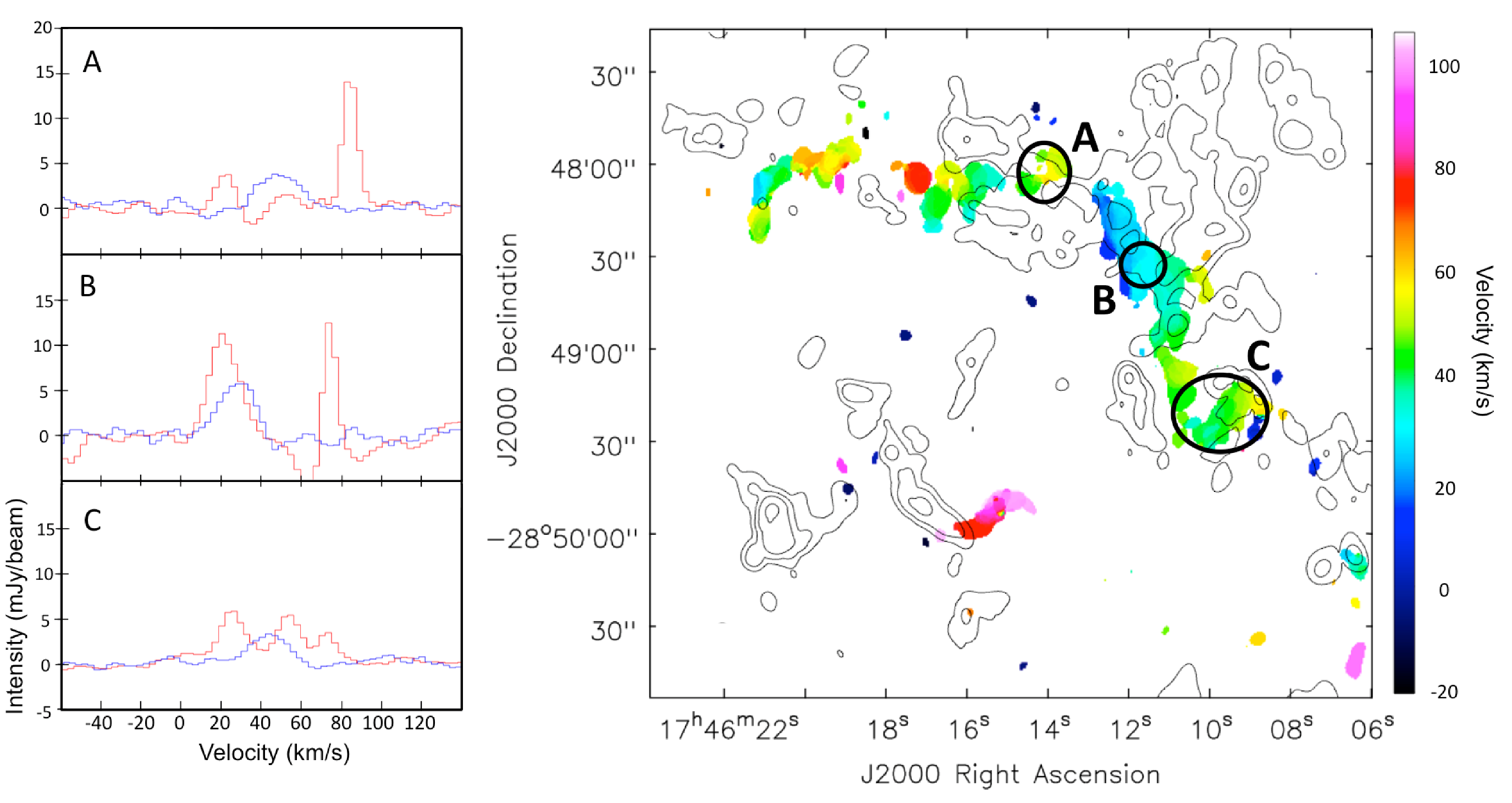}
\caption{({\it left}): spectra of both the molecular (red: \am~(3,3)) and ionized (blue: H63$\alpha$+H64$\alpha$) gas toward \sick. ({\it right}): color scale shows the intensity-weighted velocity distribution of the combined H63$\alpha$+H64$\alpha$ radio recombination lines over -20$-$100 \kms, with black contours of the \am~(3,3) emission at 5, 10, 20, \andd~50 $\times$ 3 m\jyb. The contours and color scale in {\it right} are from Figure \ref{ra mole}, top left and bottom right, respectively.  }
\label{NH3spec}
\end{figure*}

 \begin{figure}
\includegraphics[scale=0.3]{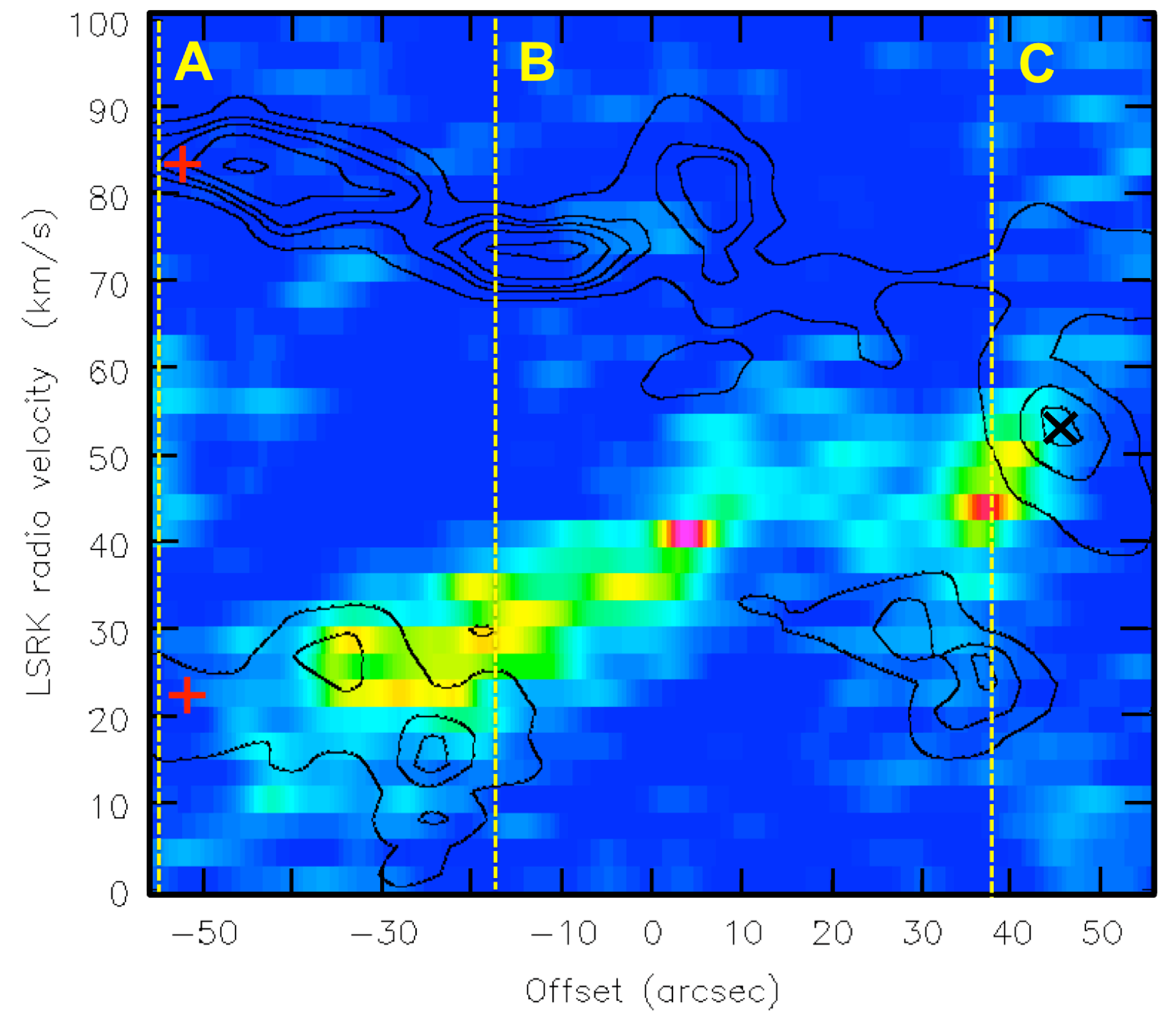} 
\caption{Position-velocity~comparison of the ionized gas (color scale: H63$\alpha$+H64$\alpha$; width=18\arcsec) and the molecular gas (black contours: \am; width=41\arcsec) from 17\h46\m12\fs9, -28\degr48\arcmin02\farcs7 ($-$50 offset) to 17\h46\m09\fs1 -28\degr49\arcmin25\farcs8 (+50 offset). 
The spectral and spatial resolution for both transitions are \til3 \kms~and \til5\arcsec, respectively (see Section \ref{obs}). The three yellow vertical dashed lines show the relative locations of the three regions in Figure \ref{NH3spec}.
The red ``$+$'' signs mark the midpoint of the ellipse discussed in Section \ref{pvsection}. The black ``$\times$'' sign indicates the edge of ellipse, also discussed in Section \ref{pvsection}. }
\label{sicklepv}
\end{figure}

\subsection{Morphology and Kinematics of Ionized Gas}
\label{recombtext}

Figure \ref{ra mole} (right) shows the ionized gas emission, as traced by the combined H64$\alpha$ and H63$\alpha$ radio recombination lines. Figure \ref{ra mole} (top right) shows that the morphology of the ionized gas emission follows the continuum emission. The brightest ionized gas emission in the \rar~($>$11 m\jyb) coincides with \pistol. Ionized gas intensities in \sick~range from 6$-$10 m\jyb. Central velocities fit to the ionized gas profiles in \sick~and Pistol nebula range from 15$-$75 \kms~and 80$-$150 \kms, respectively (Figure  \ref{ra mole}, bottom right). The velocity of the ionized gas in the Sickle \hii~pillars are typically higher (30$-$80 \kms) than velocities in the southern regions of \sick~(15$-$55 \kms). These velocities in the southern region of \sick~are increasing from $\sim$15 \kms~(at the Bend) to $\sim$55 \kms~in the direction of \sgra, for a gradient of $\sim$10 \kmsp~parallel to the Galactic plane. The ionized gas velocities in \sickp~do not appear to follow a gradient, as seen in the southern regions, but are disjointed in velocity space at our resolution. We do not detect any emission above 5.5 m\jyb~(5$\sigma$) from \slbv~or the surrounding shell. We also do not detect any ionized gas emission above 5$\sigma$ associated with the point source N3 or the \gcloud~\mc. 


\subsection{Comparison of Ionized and Molecular Gas along \sick} 
\label{ion+mol}

Figure \ref{NH3spec} (right) shows the spatial distribution of the molecular and ionized gas emission. The three selected locations, labeled A$-$C, were chosen because they contain both \scloud~molecular components and an ionized gas component. The molecular and ionized spectra for each location are shown as panels in Figure \ref{NH3spec} (left). At all three locations we detect only one ionized component and 2$-$3 molecular components. 

At \rega, the \high~molecular component is roughly three times brighter than the \low~molecular component. The ionized gas emission is located at intermediate-velocities between the two components and is not associated with a single molecular component. \Regb~shows a comparable intensity between the \loww~and \high~molecular components. However, the velocity profile of the \low~molecular component is broader (\til20 \kms) than the \high~molecular component (\til10 \kms). The \low~molecular component also correlates fairly closely with the ionized gas emission (Figure \ref{NH3spec}). 

The southernmost region, C, has the most complex molecular kinematics out these three regions. 
At \regc~there is a {\it third} molecular component (45$-$65 \kms) that is not detected in regions A \andd~B. This `intermediate-velocity' molecular component peaks in intensity around 55 \kms, similar to the molecular emission detected in \gcloud~(Figure \ref{all am}, bottom right). \Regc~is also located the closest in projection to \gcloud~out of all three regions. 
The ionized gas in \regc, overlaps in velocity space with the \loww~and intermediate-velocity molecular components. However, since the ionized gas emission peaks between these two components (at \til45 \kms), it is not solely associated with either molecular component (compared to the ionized gas at \regb). 

Figure \ref{sicklepv} compares the position and velocity distribution of the molecular and ionized emission along \sick~(from regions~A$-$C). The \high~molecular component shows continuous emission along this region that is {\it increasing} in velocity from \regc~(\til55 \kms) to \rega~(\til90 \kms). The ionized gas also shows continuous emission along the same path, but is {\it decreasing} in velocity from \regc~(\til60 \kms) to \rega~(\til15 \kms), consistent with the gradient described in Section \ref{recombtext}. The \low~molecular component is detected in small, disjointed clumps that are closely correlated in both position and velocity with the ionized gas emission. Both of these described velocity gradients, shown in Figure \ref{sicklepv}, are comparable in magnitude (\til10 \kmsp) but opposite in sign. Further, these two velocity gradients are shown to spatially overlap in both position and velocity near \regc. 


\section{Discussion}       
\label{dis}   


\subsection{Arrangement of Gas Components in \scloud}
\label{arrangement} 


 \begin{figure}
\includegraphics[scale=0.4]{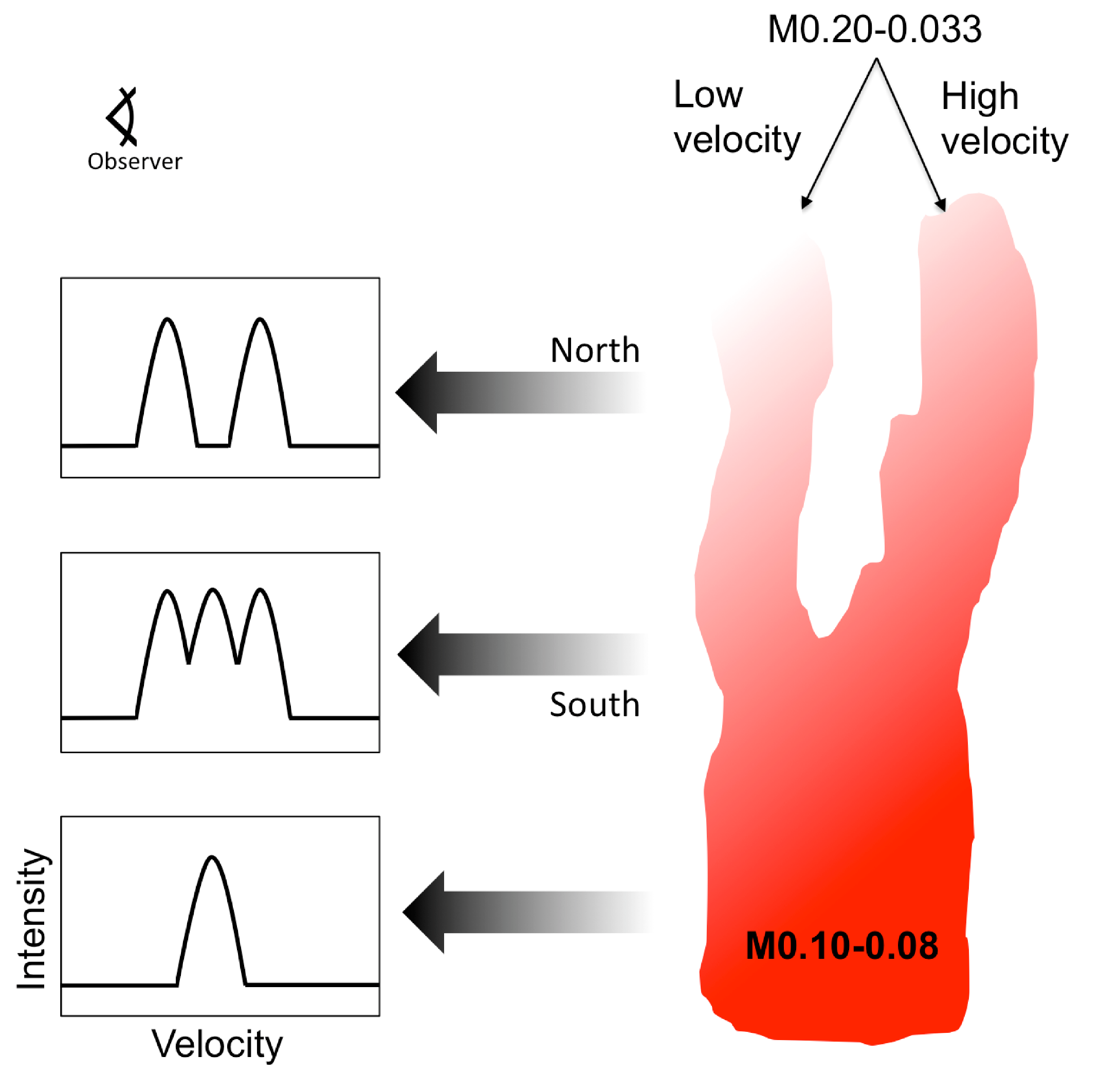}
\caption{
Schematic of the proposed possible arrangement of molecular components in \scloud~and \gcloud~(see Section \ref{arrangement}). The panels on the left show what the spectra for integrated regions in the proposed arrangement might look like, with the arrows indicating the corresponding region of the molecular gas. }
\label{cartoon}
\end{figure} 

\begin{figure*} 
\includegraphics[scale=0.64]{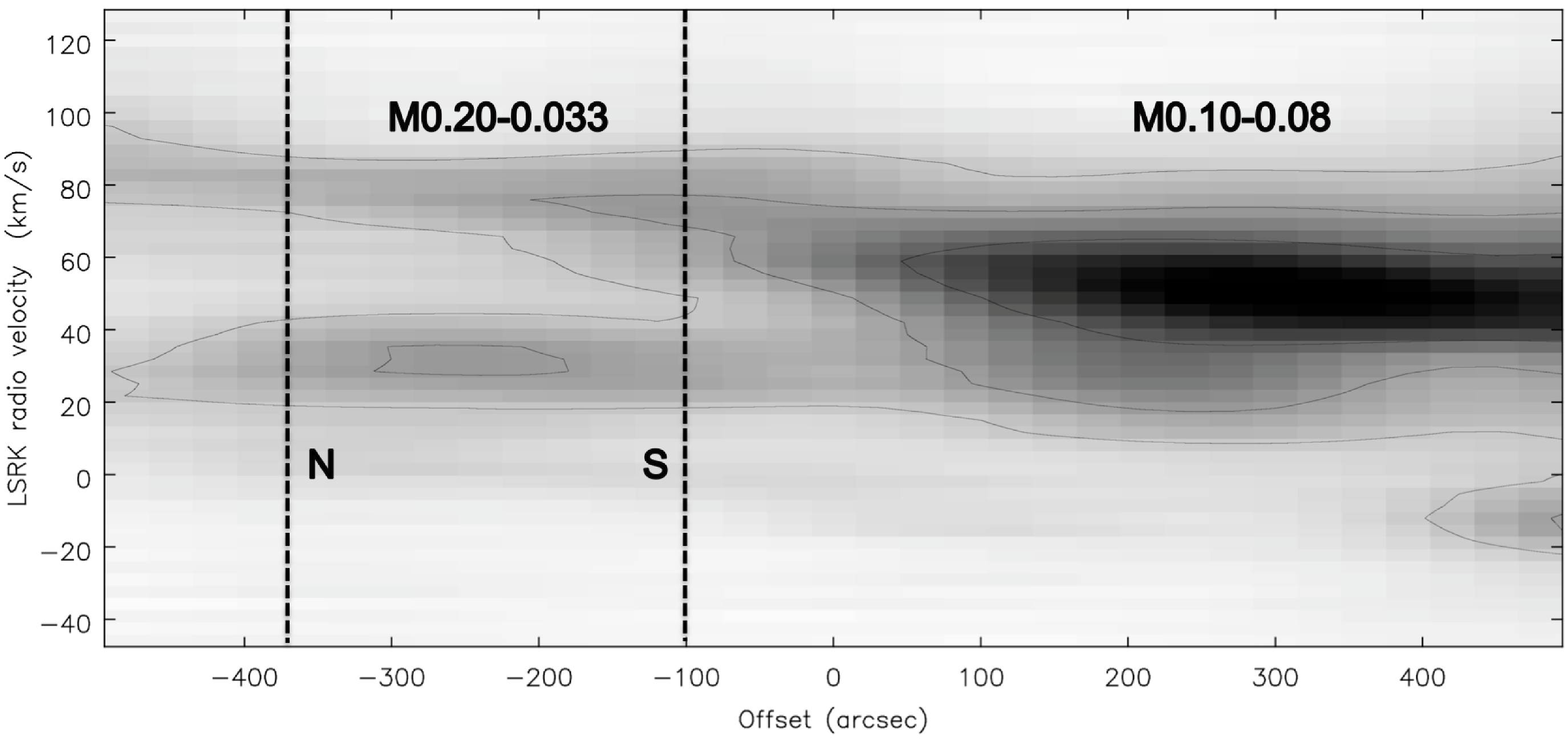} 
\caption{Position-velocity diagram of \am~(3,3) observed with the single-dish Mopra telescope as part of the HOPS survey \citep[\til2\arcmin~resolution;][]{Walsh11a}. The emission covers a galactic longitude from 0$\fdg$025~(+450\arcsec~offset) to 0$\fdg$3~($-$450\arcsec~offset), and is integrated over a 0$\fdg$15~wide slice, at a latitude of $b=-$0$\fdg$075. The vertical dashed lines represent the north (N) and south (S) regions of the cloud, as illustrated in Figure \ref{cartoon}.}
\label{SDfig}
\end{figure*}

\subsubsection{Physical Connection Between \Loww~and \High~Components} 
\label{connection}

Figure \ref{sicklepv} shows two velocity gradients: a gradient {\it increasing} from C to A (\til55$-$90 \kms; the \high~component) and a gradient {\it decreasing} from C to A (\til60$-$15 \kms; the \low~component). 
These two velocity gradients spatially overlap in \pv~space at an offset of \til45\arcsec~in Figure \ref{sicklepv} (\ie, near C). At this location we detect a faint bridge of continuous emission ranging in velocity from \til35$-$80 \kms~over a distance of \til10\arcsec. This large range of velocities is consistent with the spectrum of the molecular and ionized gas at \regc~in Figure \ref{NH3spec}. This spectrum shows continuous emission via a third intermediate-velocity molecular component around 55 \kms, which bridges the \loww~and \high~molecular components. 
Therefore, the bridge in \pv~space suggests the two velocity components are physically {\it connected} near \regc. 
The absence of this intermediate-velocity molecular component at regions A \andd~B indicates the two velocity components are physically {\it separated} at these locations. 

Figure \ref{cartoon} shows a schematic of the proposed possible arrangement of molecular components in M0.20$-$0.033 and \gcloud~(see Section \ref{arrangement}). The panels on the left show what the spectra for integrated regions in the proposed arrangement might look like, with the arrows indicating the corresponding region of the molecular gas based on data presented in Figure \ref{NH3spec} and \ref{sicklepv}. For example, the profiles for Regions A and B in Figure \ref{NH3spec} are illustrated by the `North' \scloud~profile at the top of Figure \ref{cartoon}. The proposed spectrum of molecular emission for Region C in Figure \ref{NH3spec} is illustrated by the `South' \scloud~profile in Figure \ref{cartoon}, and the proposed spectrum for \gcloud~is represented by the lower profile in Figure \ref{cartoon}. 

In the north of \scloud, our simple interpretation is that these molecular components may lie at physically different distances but are part of the same molecular structure. In contrast, our simple interpretation of the triple-peaked profile in the south of \scloud~is that the profile represents a superposition of the two molecular components from \scloud~as well as the molecular gas at an intermediate velocity, possibly associated with \gcloud. However, the sight line to the \gc~is incredibly complex, and we acknowledge that our interpretation may be oversimplified. The discussion that follows is an attempt to quantify and characterize the arrangement in more detail. 

However, it is important to note that we also have kinematic information of the ionized gas as well. Region A in Figure \ref{NH3spec} shows two molecular components (as depicted in the north of Figure \ref{cartoon}) with an intermediate-velocity ionized gas component. The combination of these three components (two molecular and one ionized) results in continuous emission, from \til15 to 90 \kms, in the northern region of \scloud. This bridging of the two molecular components by an ionized gas component may be an indication that the emission in \scloud~is also physically connected in the north. This possible connection in the north, via an intermediate-velocity ionized gas component, is consistent with the gas distribution shown in Figure \ref{ra mole}. For example, the bright clump of high-velocity ionized gas emission in the \sickp~(\til80 \kms; 17\h46\m17\fs5, $-$28\degr48\arcmin10\arcsec) is spatially adjacent to the large structure of high-velocity molecular emission evident in Figures \ref{ra mole} \& \ref{NH3spec}, indicating that this feature is likely to be more closely related to the high-velocity molecular component than the low-velocity molecular component. A more detailed discussion of the ionized gas emission is presented in Section \ref{ion-dis}. 


\subsubsection{The Orbital Stream: Connection of \scloud~to Nearby Molecular Clouds}
\label{orbit}

The molecular spectrum of \gcloud~shows a single velocity component, \til55 \kms, located between the \loww~and \high~components of \scloud~(Figure \ref{all am}, right). The southern region of \scloud~is closest in projection to \gcloud, suggesting the intermediate-velocity spectral feature detected toward \regc~might be an extension of emission from \gcloud. Figure \ref{cartoon} further illustrates a possible relationship of the two velocity components of \scloud~to \gcloud. In this schematic both velocity components of \scloud~are physically connected to \gcloud. 

The connection between \gcloud~and the \high~molecular component in \scloud~is also posited in the \cite{Kru15} orbital model, which places both on the near side of the \gc~(stream 1).  The \low~molecular component of \scloud, however, is proposed to be located on the far side of the \gc~in the \cite{Kru15} orbital model (stream 3). However, the data presented here and the single-dish observations of \cite{SG91} indicate that the two molecular components in \scloud~are spatially connected as they show continuous emission in \pv~space (Section \ref{connection}).

\begin{figure*}
\includegraphics[scale=0.64]{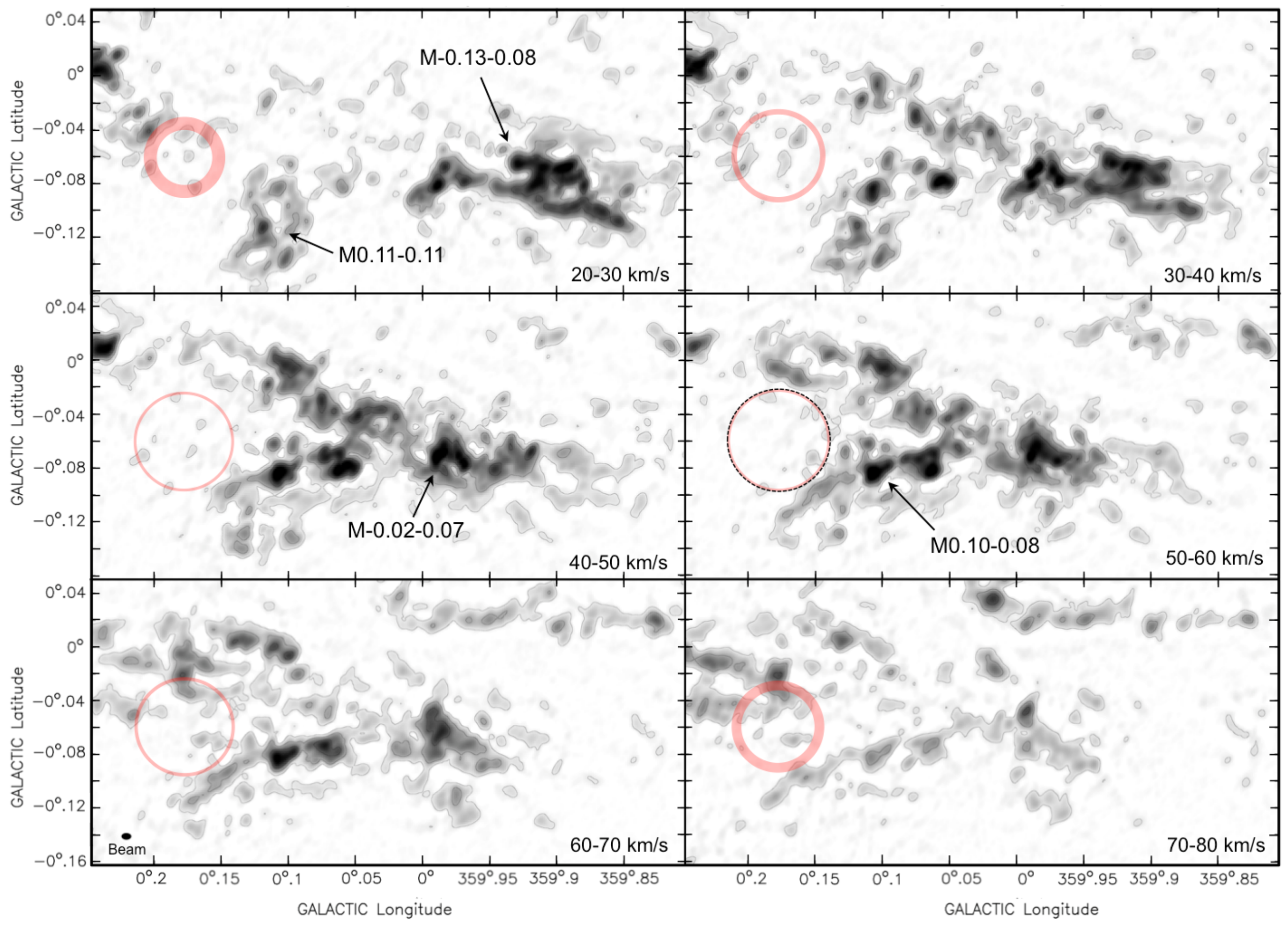}
\caption{Maximum intensity \am~(3,3) emission from ATCA \citep[SWAG survey; angular resolution of \til20\arcsec~(bottom left corner);][]{ott16} integrated over $\Delta$v=10 \kms~intervals,~as indicated in the bottom right corners. The black contours show: 5, 25, \& 75 $\times$ 15 m\jyb~(average rms noise over all panels). The black dashed circle shows the size of the expanding shell at the systemic velocity (53 \kms; see Section \ref{pvsection}). The red shaded annuli represented the predicted sizes of the shell over different velocity ranges and a spherical geometry (see Appendix \ref{los-geo}). }
\label{atca-panels}
\end{figure*}

In order to further investigate the proposed connection between \scloud~and nearby clouds~in the region, we incorporate observations made with the Mopra single-dish telescope.\footnote{Single-dish \am~(3,3) observations from the HOPS survey \citep[\water~southern Galactic Plane Survey;][]{Walsh08, Walsh11a, Walsh11, purcell12}, with \til2\arcmin~resolution.} Figure \ref{SDfig} shows the \pv~distribution of the large-scale cloud structures from \scloud~to \gcloud.\footnote{Note that all figures with a relatively large field of view ($>$0\fdg25; Figures \ref{SDfig}, \ref{atca-panels}, \& \ref{bubble}) are presented in galactic coordinates. All figures with a smaller field of view ($<$0\fdg25; Figures \ref{sicklecont}$-$\ref{sicklepv} \& \ref{sickle-shell-finder}) are presented in equatorial coordinates.} 
This \pv~diagram covers a larger and wider spatial range than \cite{SG91}, who focused on the region around \scloud~(see their Figures 2 \& 4). Figure \ref{SDfig}~supports the interpretation that the two velocity components in \scloud~are physically connected in the south near \gcloud. Figure \ref{SDfig}~also shows that the molecular gas appears to be physically connected between \scloud~ and \gcloud, further supporting the hypothesis that the two clouds reside along the same \gc~stream \citep[stream 1 of][]{Kru15}. 
While these single-dish Mopra observations (Figure \ref{SDfig}) are consistent with the kinematic structure inferred from our VLA observations (Figure \ref{sicklepv}), many of the small-scale features ($<$1\arcmin$-$2\arcmin) are not detected at the resolution of MOPRA. Therefore, we turn to a survey having intermediate resolution, and which therefore shows the relationship between large (\til1\arcmin$-$2\arcmin) and small (\til10\arcsec$-$20\arcsec) scale structures.  

Figure \ref{atca-panels} shows the \am~(3,3) distribution from the ATCA telescope,\footnote{As part of the SWAG survey (Survey of Water and Ammonia in the \gc), \til20\arcsec~spatial resolution, 2 \kms~spectral resolution. See \cite{ott16} for further details.} over a velocity range of \til20$-$80 \kms. This velocity coverage~includes both the low- and high-velocity molecular components of \scloud. The molecular emission generally appears to be concentrated towards the eastern Galactic side of the observed region at higher velocities (\eg, v $>$ 60 \kms). The brighter molecular emission at low velocities (\eg, v $<$ 30 \kms) is mainly concentrated towards the western Galactic side (\eg, \twenty~cloud). As \gc~clouds are known to have large velocity dispersions \citep[\til20$-$50 \kms;][]{Bally87} individual clouds can span multiple velocity panels, as seen in Figure \ref{atca-panels}. However, there does appear to be a general trend of increasing velocity in the direction away from \sgra~(\ie, west to east). Assuming a $\Delta$v of \til60 \kms~over a distance of 0\fdg3, implies a velocity gradient of \til1.4 \kmsp,~which is comparable to the slope of stream 1 (\til1 \kmsp) in the \cite{Kru15} orbital model.
This large-scale velocity gradient is an order of magnitude smaller than the velocity gradients observed in both molecular components of \scloud~(\til10 \kmsp; see Figure \ref{sicklepv} and Section \ref{ion+mol}). Therefore, the observed velocity gradients in \scloud~appear to be the result of localized kinematics in the cloud rather than the large scale motion of the orbital stream.


\subsection{\scloud: an Expanding Shell?}
\label{swag shell}

A plausible way to describe the separation in the northern part of \scloud~is with an expanding shell. 
In this scenario the \loww~and \high~molecular components correspond to gas on opposite sides of the shell. The resulting Doppler shift from this scenario produces a blueshifted velocity component on the near side of the shell (\til 25 \kms; \low~molecular component).
The far side of the shell would have a redshifted velocity component (\til 80 \kms; \high~molecular component). 
The systemic velocity of the shell would have a velocity centered roughly between the two values (\ie, \til50 \kms). Emission detected at the systemic velocity would correspond to the edge of the shell (\ie, where the two velocity components are connected), since the gas at this location is expanding in a direction perpendicular to our \los. 

The estimated systemic velocity is comparable to the intermediate-velocity component in \regc~(\til55 \kms, Figure \ref{NH3spec}), which bridges the velocity range between the \loww~and \high~molecular components. \Regc~is also the location where we argue that the two velocity components are physically connected (see Section \ref{connection}). The absence of \til50 \kms~molecular emission in \regs~A \andd~B (Figure \ref{sicklepv}) is consistent with the gas having been Doppler shifted away from the systemic velocity to higher and lower velocities (two sides of the shell). 
The suggested systemic velocity is also consistent with the velocities of other \mcs~located on the same orbital stream (\eg, \gcloud) as argued in Section \ref{orbit}.  

\hi~observations by \cite{lang10} support our arrangement of components in \scloud~(see \los~placement in Figure \ref{cartoon}). 
The 21 cm line of atomic hydrogen (\hi), is found in the intermediary region between \hii~regions and molecular gas (\eg, \htwo, \am). Therefore, detecting (or not detecting) \hi~absorption features is useful in the \los~arrangement of components, with respect to bright continuum regions.  
Across \sick, \hi~absorption is detected at velocities associated with the \low~molecular component of \scloud. This implies that some part of the \scloud~\low~molecular component~is located {\it in front} of \sick, along our \los. No \hi~absorption is detected at velocities consistent with the \high~molecular component \scloud, indicating this component is {\it behind} \sick, and thus {\it behind} the \low~molecular component, along our \los.  
We note that the absorption-line data are therefore inconsistent with the \cite{Kru15} scenario in which the low-velocity component is located on the far sides of the GC and therefore behind the high-velocity component.


\subsubsection{Morphological Evidence for an Expanding Shell} 
\label{shell-text}

For a uniformly expanding shell, one would expect to see a circular ring of emission at the systemic velocity,~corresponding to gas along the edge of the shell (see Section \ref{swag shell}).
The \scloud~expanding shell appears to have an angular extent of a few arcminutes~(based on Figure \ref{sicklepv}), and a systemic velocity around 50 \kms~(see Section \ref{swag shell}). 
The 50$-$60 \kms~panel in Figure \ref{atca-panels} shows a cavity in the molecular emission that is open on the eastern Galactic side. 
This \til50 \kms~cavity in the molecular gas can be described by an expanding shell that is 
centered at \ra=17\h46\m17\s, \dec=$-$28\degr49\arcmin00\arcsec, with radius $r$=135\arcsec~(5.2 pc); consistent with the predicted values. 
However, the apparent radial size of an expanding shell changes as a function of the observed Doppler-shifted velocity (see Appendix \ref{los-geo}). The red annuli in Figure \ref{atca-panels} show the predicted angular size of the shell, at the indicated velocities, using the technique shown in Figure \ref{radial-size} and the expansion velocity derived in Section \ref{pvsection}.

These ATCA observations show brighter molecular emission ridges along the western Galactic side of the dashed circle~at 50$-$60 \kms, with little molecular emission detected towards the center and eastern Galactic side (Figure \ref{atca-panels}). 
Clumps of molecular emission are detected along the shaded red annuli in all six panels. For example, several clumps in the 20$-$30 \kms~panel, the \low~molecular component, correspond well with the predicted extent of the proposed shell. This correlation between the predicted distribution of gas and the detected large scale structure supports the idea that the \low~component is associated with the expanding shell. Although the detected molecular emission is not constrained to the region inside the shaded red annuli, we note that the schematic in Figure \ref{radial-size} assumes a thin shell ($\Delta$r$<<$r) and isotropic expansion. Thus deviations in the expansion velocity, from possible density fluctuations, could distort the ring-like structure predicted for an expanding shell.

The lack of molecular emission towards the eastern Galactic side of the proposed expanding shell, at 50-60 \kms, is noticeable. 
However, molecular emission is detected towards the north-eastern Galactic side of the red annulus at slightly higher velocities (see the 60$-$70 \kms~panel in Figure \ref{atca-panels}). 
Additionally, the 40$-$50 \kms~panel also shows several clumps of emission along the eastern edge of the red annulus with relatively low emission (\til5$\sigma$) inside the region. These detections may be an indication that the systemic velocity of the proposed \mshell~extends over a range of velocities within 15 \kms~of \til55 \kms).

The physical size of the described molecular cavity in M0.20$-$0.033 
is consistent with that of other molecular shells in the \gc~\citep[2$-$8 pc;][]{Tsuboi97, Oka01c, Tsuboi09}. The edges along this cavity in the molecular gas also coincide with ridges of brighter dust emission detected in the Herschel SPIRE 160, 250, 350, and 500 $\mu$m bands \citep[see the 250 $\mu$m emission in ][their Figure 2]{Molinari11}. The correlation between the dust ridges and molecular gas implies that the proposed expanding shell could also be influencing the dust in the region. Although the morphology of the molecular gas in the 50$-$60 \kms~range does suggest an expanding shell, detailed analysis of \pv~slices can help constrain the estimated values of the expansion and systemic velocities for the proposed expanding shell. 

\begin{figure}
\includegraphics[scale=0.39]{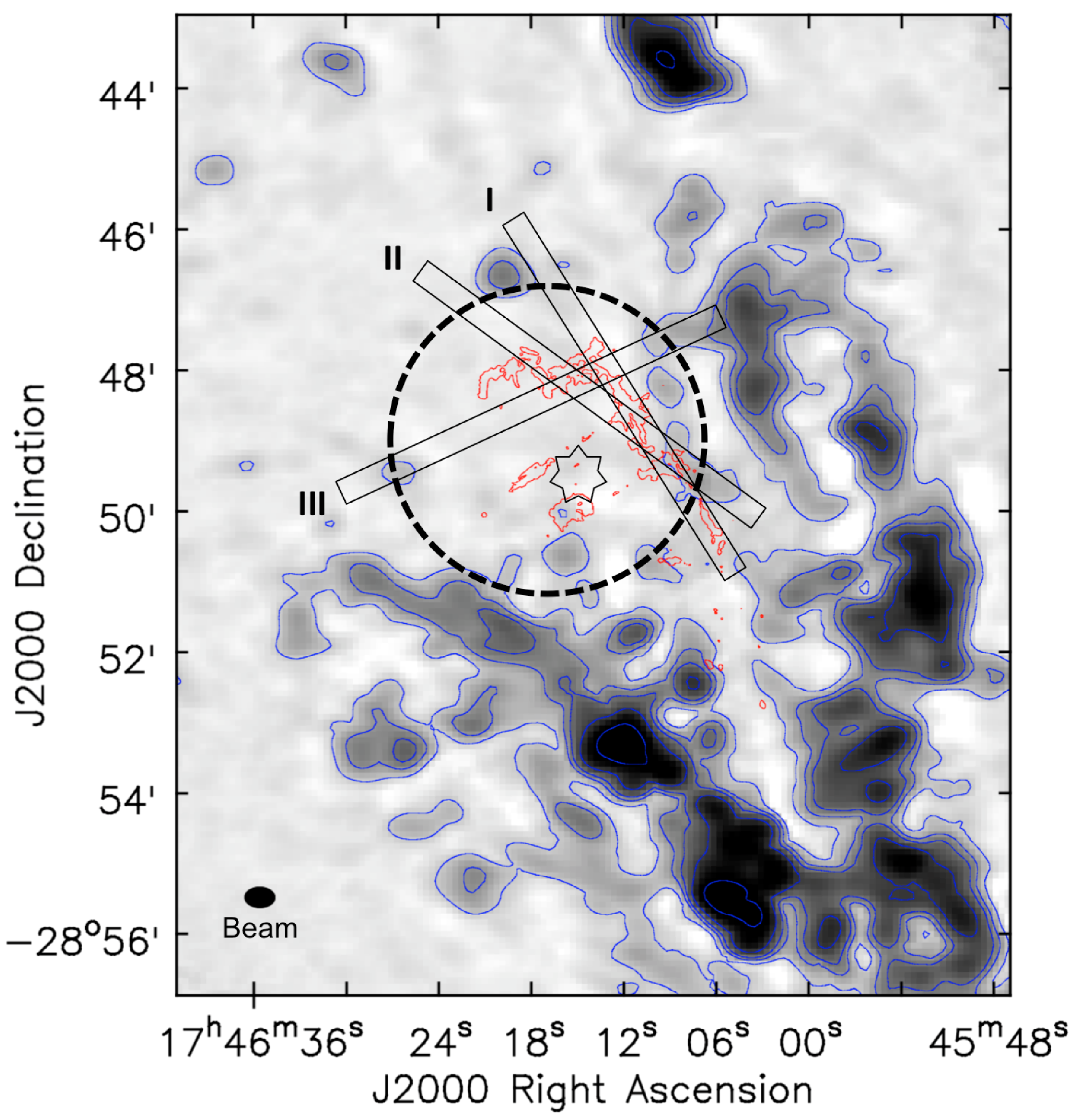}
\caption{Maximum intensity \am~(3,3) emission from ATCA \citep[SWAG survey; \til20\arcsec~resolution, bottom left corner;][]{ott16} over a velocity range of 45$-$65 \kms, with blue contours showing 5, 15, 25, 50 \andd~125 $\times$ 20 m\jyb~(rms level). The red contour represents the 24.5 GHz continuum emission (from Figure \ref{sicklecont}) above 0.5 m\jyb~(10$\sigma$ level). The black dashed circle shows the proposed size and location of the \mshell~(see Figure \ref{atca-panels} and Section \ref{shell-text}). The star shows the location of the \qc. The black boxes show the relative lengths (350\arcsec) and integrated widths (5\arcsec) of~slices I$-$III in Figure \ref{sickle-shell-pv}. } 
\label{sickle-shell-finder}
\end{figure}


\subsubsection{Kinematic Properties of the \scloud~Expanding Shell} 
\label{pvsection}

Using the geometry of an expanding shell, we can estimate the systemic velocity, \vsys,~and expansion velocity, \vexp, (see Appendix \ref{app} for more details). We will assume the gas distribution in Figure \ref{sicklepv} delineates the right half of the ellipse (\ie, semi-major axis in Figure \ref{shell-model}, right). Therefore, we identify the dashed line at location `A,' in Figure \ref{sicklepv}, as the midpoint of the chord. Thus, the semi-major axis of this ellipse (\ie, r$\cdot$cos$\theta$, where r=135\arcsec) has an angular length of \til100\arcsec. We measure a minimum velocity (\vmin) of 23$\pm$3 \kms~and a maximum velocity (\vmax) of 83$\pm$3 \kms~at the midpoint of this chord, resulting \vsys~equal to 53$\pm$3 \kms. This \vsys~corresponds well to the velocity measured at the edge of the shell (52$\pm$3 \kms) in Figure \ref{sicklepv}. Using \vmin~and \vmax~for this chord, located at an angle $\theta$ of 42\degr~from the origin, we estimate \vexp=40$\pm$3 \kms.

Next, we can look for consistency between predicted values for \vmin~and \vmax~and the values of the \vmin~and \vmax~velocities measured in the molecular gas on a slightly larger scale. We can compare predicted values for \vmin~and \vmax~(using \vsys~and \vexp~as above) with the distribution of velocities in the \pv~diagram for the lower resolution (\til2\arcmin~resolution) molecular gas data shown in Figure \ref{SDfig}. The predicted values for \vmin~range from 20$-$35 \kms~and for \vmax~range from 75$-$90 \kms~for a range of angles, $\theta$, 30$-$60\degr. The \vmin~and \vmax~measured in Figure \ref{SDfig} along the northern region of the cloud (`N') are 20$-$40 \kms~and~75$-$85 \kms, respectively. These values are in good agreement with the predicted values, illustrating that the parameters we suggest for the expanding shell are consistent with the data. 

\begin{figure}
\includegraphics[scale=0.58]{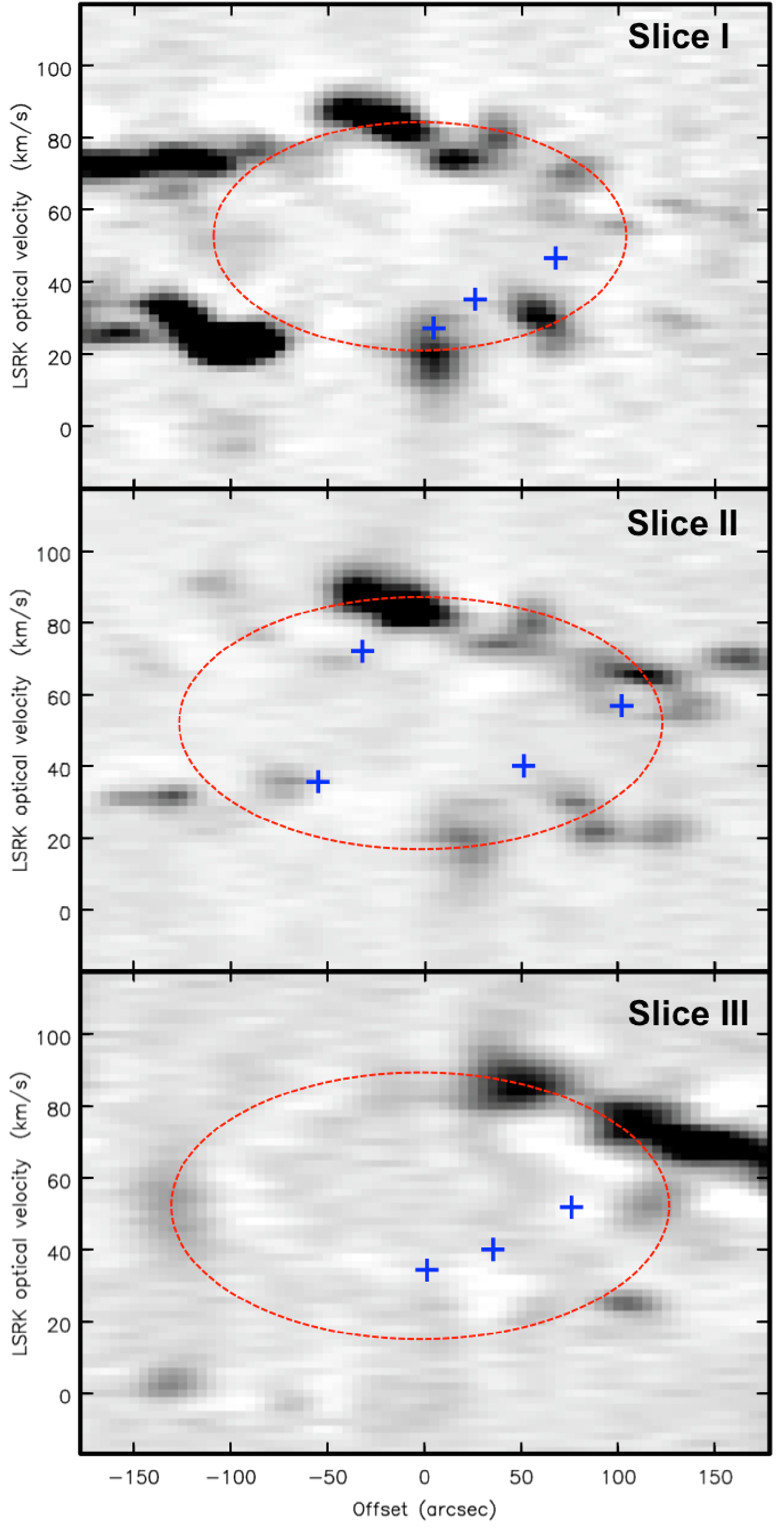}
\caption{Position-velocity~distribution of \am~(3,3) from ATCA of slices I$-$III (Figure \ref{sickle-shell-finder}). The red dashed ellipse shows the derived \pv~distribution from Figure \ref{shell-model} using a \vsys~of 53 \kms~and a \vexp~of 40 \kms~from Section \ref{shell-text}. The chord length of the three different slices varies from 229\arcsec~(Slice I) to 255\arcsec~(Slice III). The blue crosses show the central velocity emission of the ionized gas at several locations across the slice (Section \ref{ion-dis}). } 
\label{sickle-shell-pv}
\end{figure}

In order to further examine the \pv~structure expected from an expanding shell (Figure \ref{shell-model}, right), we examine the ATCA data, as its angular resolution of \til20\arcsec~provides a compromise between the VLA data (5\arcsec~resolution) and the Mopra data (\til2\arcmin~resolution). Figure \ref{sickle-shell-finder} shows the 45$-
$65 \kms~molecular emission (\ie, roughly $\pm$10 \kms~from the suggested 53 \kms~systemic velocity, see discussion in Section \ref{shell-text}), with the relative locations of: the \mshell, \sick, and the \qc.
We performed a \pv~analysis on three slices (labeled slice I$-$III) across the \mshell~as indicated by the black boxes in Figure \ref{sickle-shell-finder}. The three panels in Figure \ref{sickle-shell-pv} (grayscale) show the \pv~distribution of the observed molecular emission for each slice. These three panels were created using the CASA task IMPV, with a fixed length of 350\arcsec~and the center location of the chord. We measured the chord length for each slice and inferred the corresponding \vmin~and \vmax~of the ellipse, using the derived \vsys~and \vexp~from above. The dashed lines in Figure \ref{sickle-shell-pv} show the resulting predicted ellipse, with the end points ($\pm$r$\cdot$cos$\theta$) corresponding to the edge of the shell (outlined in Figures \ref{atca-panels} \& \ref{sickle-shell-finder}).\footnote{Note that because the location of these end points varies (Figure \ref{sickle-shell-finder}), the chord length of each slice also varies (\ie, slice I is shorter than slice III), as seen in Figure \ref{sickle-shell-pv}.} All three predicted ellipses generally follow the observed \pv~distribution, indicating that our estimated systemic and expansion velocities are consistent with the observed emission.\footnote{Our schematic in Figure \ref{shell-model} assumes a uniformly expanding shell, however, variations in gas density can effect the expansion velocity. These variations in the gas density can cause perturbations from the predicted elliptical distribution in \pv~space.} We conclude that the \pv~structure supporting an expanding shell in the \mc~\scloud~is present in all of our observations that represent a range of spatial scales (from the high-resolution VLA data to the single-dish Mopra data).


\subsubsection{Age and Orbital Stream of the M0.20$-$0.033 Expanding Shell} 
\label{shell-compare}

Using the assumption that the time derivative of \vexp~is zero, we can estimate an upper limit on the age of the shell (\tage=1.3$\times$10$^{5}$ yr). The age and expansion velocity (40 \kms) of the \mshell~are comparable with properties of other molecular shells in the \gc~\citep[0.7$-$1.5$\times$10$^5$ yr, and \til20$-$30 \kms; ][]{Tsuboi97, Oka01c, Tsuboi09}. These molecular shells are all located within the inner 100 pc of the \gc~and are thought to be the result of several supernovae or a few hypernovae. 

The central velocities detected in M0.10$-$0.08 (Section \ref{orbit}) are consistent with the derived systemic velocity of the \mshell~(53 \kms). 
The consistency between the systemic velocities of these two clouds further supports our argument that these clouds are physically connected. 
If these two clouds are physically connected (\ie, on the same orbital stream) then the Doppler shifted velocities detected in \scloud~might be the result of a perturbation by the expanding shell. 
The perturbation in this stream might be the result of the \qc,~which is \til30\arcsec~(1.2 pc in projection) from the center of the \mshell~(Figure \ref{sickle-shell-finder}). The expansion velocity of the \mshell~is also consistent with velocities from wind-blown bubbles around massive clusters \citep[\eg,][]{Dent09}, indicating that this massive cluster might be the driving force behind the expanding shell. 

The proposed interaction between the \qc~and the expanding shell is consistent with the location of the cluster in the \cite{Kru15} orbital model. Based on the velocity of the \qc~\citep[\til90$-$100 \kms;][]{Stolte14}, \cite{Kru15} suggest that the cluster is also located on the near side of the \gc~(stream 1). In addition, our observations help to support the conclusions of \cite{Stolte14}, who argue that the parent cloud of the \qc~is not on a circular orbit around the \gc, but rather a non-circular orbit, such as the \cite{Kru15} orbital streams.  


\subsection{The \qc: Driving Force Behind the Expanding Shell?}
\label{ion+hi}

The massive Quintuplet star cluster~\citep[\til10$^{50.9}$ photons s$^{-1}$, age: 4.8$\pm$1.1 Myr;][]{figer99a, schneider14} has a large population of massive stars (OB supergiants) and several unusual types of Wolf-Rayet stars. In total, this cluster is capable of ionizing the edge of the nearby (\til1\arcmin; 2.3 pc in projection) \scloud~\mc, producing \sick~\citep[2.8$\times$10$^{49}$ photons s$^{-1}$, 240 \msun;][]{lang97}. In addition, radio detections of a number of the most powerful stellar winds from this cluster have been detected by \cite{lang05} and the collective winds are likely to be influencing the surrounding ISM. In order to determine whether the cluster is capable of driving the expansion, we compare the radiative momentum of the cluster (\prad) to the momentum of the expanding shell (p$_{shell}$). Since the radiative and wind energy of the cluster can be lost by cooling processes and dissipated in shocks the conserved momentum, which cannot be dissipated, provides the most informative comparison. 

Stellar clusters inject momentum into the surrounding ISM through a combination of radiation and winds (which are radiation driven). Radiative momentum from a cluster is injected into the surrounding ISM at a rate of \pdot\til$\epsilon$L$_{bol}$/c, where $\epsilon$ is the fraction of the radiation absorbed by the ISM (or covering factor; \ie, $\epsilon$$\leq$1). We assume $\epsilon$=0.5 based on the distribution of molecular gas in Figure \ref{atca-panels}, which shows an open cavity in the emission.
We will also assume the bolometric luminosity of the \qc~\citep[L$_{bol}$=10$^8$ \lsun;][]{lang05} is constant over the lifetime of the expanding shell: \tage. Therefore, the total momentum (\prad=$\epsilon$\pdot\tage) injected into the ISM by the \qc~is 6.6$\times$10$^{4}$ \msun~\kms.

Estimations of the molecular gas mass in M0.20$-$0.033 range from 0.6 to 1.3$\times10^{5}$ \msun~\citep{SG91, Tsuboi11}. However, both of these mass estimates only account for the \low~molecular component. Therefore, the total mass of \scloud~would be twice these estimates, assuming the high-velocity molecular component has roughly the same amount of molecular gas. However, the area over which this mass is derived is approximately twice the size of the shell. Thus, we will assume the mass of the shell is M$\simeq$1$\times$10$^5$ \msun, for the momentum calculation, with a $v$=\vexp. Therefore, the momentum of the expanding shell is \til4$\times$10$^{6}$ \msun~\kms.

\begin{figure*}
\centering
\includegraphics[scale=0.8]{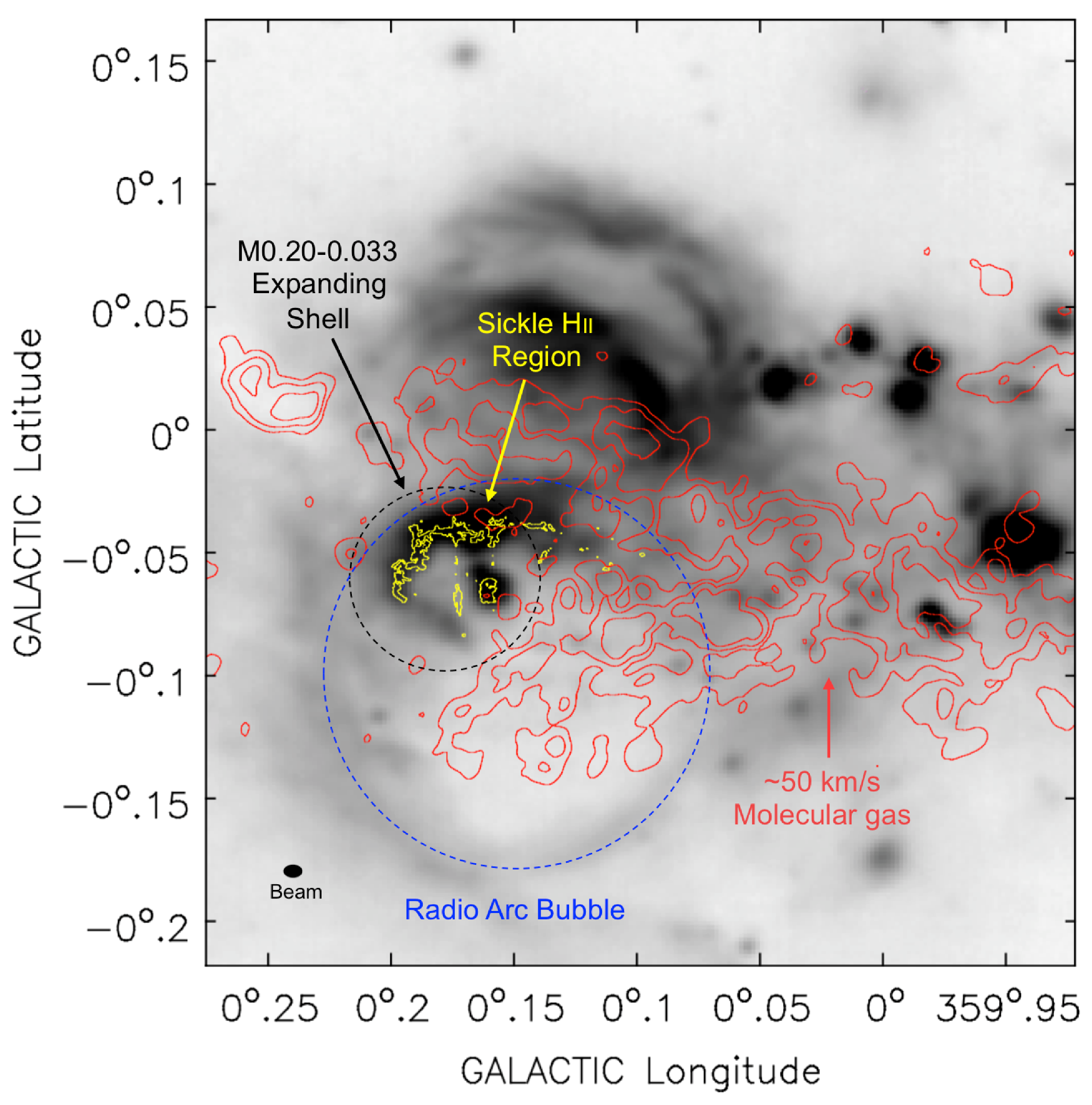}
\caption{21 $\mu$m MSX emission (Band E; greyscale, 18\farcs3 resolution) from \cite{Price01}. The blue dashed circle outlines the \cite{simpson07} \irbubble~(see their Figure 1, right). The black dashed circle shows the \mshell~(see Section \ref{shell-text}). The red contours represent the 45$-$65 \kms~molecular emission at 5, 20 \& 80 $\times$ 20 m\jyb~(from Figure \ref{sickle-shell-finder}; \til20\arcsec~resolution, bottom left corner). The yellow contour shows the 10$\sigma$ level of the 24.5 GHz radio continuum (Figure \ref{sicklecont}). } 
\label{bubble}
\end{figure*}

Clearly, the \qc~alone does not have enough radiative momentum to drive the expansion. An additional source of momentum is needed to produce the observed expanding shell. This additional momentum could be in the form of supernova explosions. Given the age of the \qc~(4.8$\pm$1.1 Myr) and the large number of evolved massive stars \citep{figer99a}, there is a high probability that some stars in the cluster have undergone a supernova explosion.  The momentum injected by a single supernova into the surrounding ISM is \psn=$\epsilon$M$_{sn}v_{sn}$, where M$_{sn}$ is the total ejecta mass and $v_{sn}$ is the initial blast wave expansion velocity. We will again assume a covering factor of $\epsilon$=0.5, since the molecular emission in Figure \ref{atca-panels} shows an half-open cavity. 
Thus, \psn\til1.3$\times$10$^5$ \msun~\kms~assuming a $v_{sn}$\til10$^4$ \kms~and M$_{sn}$\til25 \msun. 
Therefore, approximately 30 supernovae (or a few hypernovae, M$_{sn}$$>$100 \msun) are needed to produce the expanding molecular shell detected. The production of this molecular shell by a number of supernova explosions would be reminiscent of the hypothesized production of other molecular shells detected in the \gc~\citep{Tsuboi97, Oka01c, Tsuboi09}.


\subsubsection{Connection to the Radio Arc Bubble?}
\label{gcsb}

The connection between the \qc~and the molecular shell echoes several previous studies of this region. The \gc~`Radio Arc bubble' \citep{simpson07} is a large (d\til10\arcmin; 20 pc) infrared bubble, that is suggested to be due to the outflow of matter and energy from the \qc. Figure \ref{bubble} shows the relative location of the \mshell~compared to the \irbubble~presented in \cite{simpson07}. The north and northeastern Galactic rim of the \irbubble~coincide with the north and northeastern Galactic rim of the \mshell. Additionally, there is relatively bright 21 $\mu$m emission located \til1\arcmin~inside the eastern Galactic rim of the \mshell. 
The southwest Galactic side of the \irbubble, however, extends \til6\arcmin~(\til15 pc) past the edge of the \mshell~(see Figure \ref{bubble}) and is much fainter than~emission detected near the \mshell. 

We propose that the southwest side of the \irbubble~corresponds to directions in which matter and energy flowing out of the cluster have encountered the least resistance (\ie, a relatively low-density ISM) and have broken out of the smaller region confined by molecular clouds. 
The circular structure of the \irbubble~is well preserved towards the southern and western sides \citep[][their Figure 1]{simpson07} indicating a weak interaction with the ambient ISM. This idea of a larger evacuated region has been evoked in the literature, as blown out (or broken) bubbles are quite common in the Galaxy and are detected in 38\% of the 322 bubbles sampled by \cite{churchwell06}. 
The velocities measured along this southwest rim of the \irbubble~are consistent with velocities in the \low~molecular component \citep[22$-$26 \kms, positions 10$-$12 in][]{simpson07}. Therefore, if the \irbubble~is associated with the \mshell,~then the bubble emission is predicted to be located on the near side of the region. 
 Additionally, this breakout hypothesis might also account for the lack of emission in the \low~molecular component along regions of our shell (\eg, slice III in Figure \ref{sickle-shell-pv}) and in the single-dish observations by \cite{SG91}. 

\begin{figure*}
\includegraphics[scale=0.64]{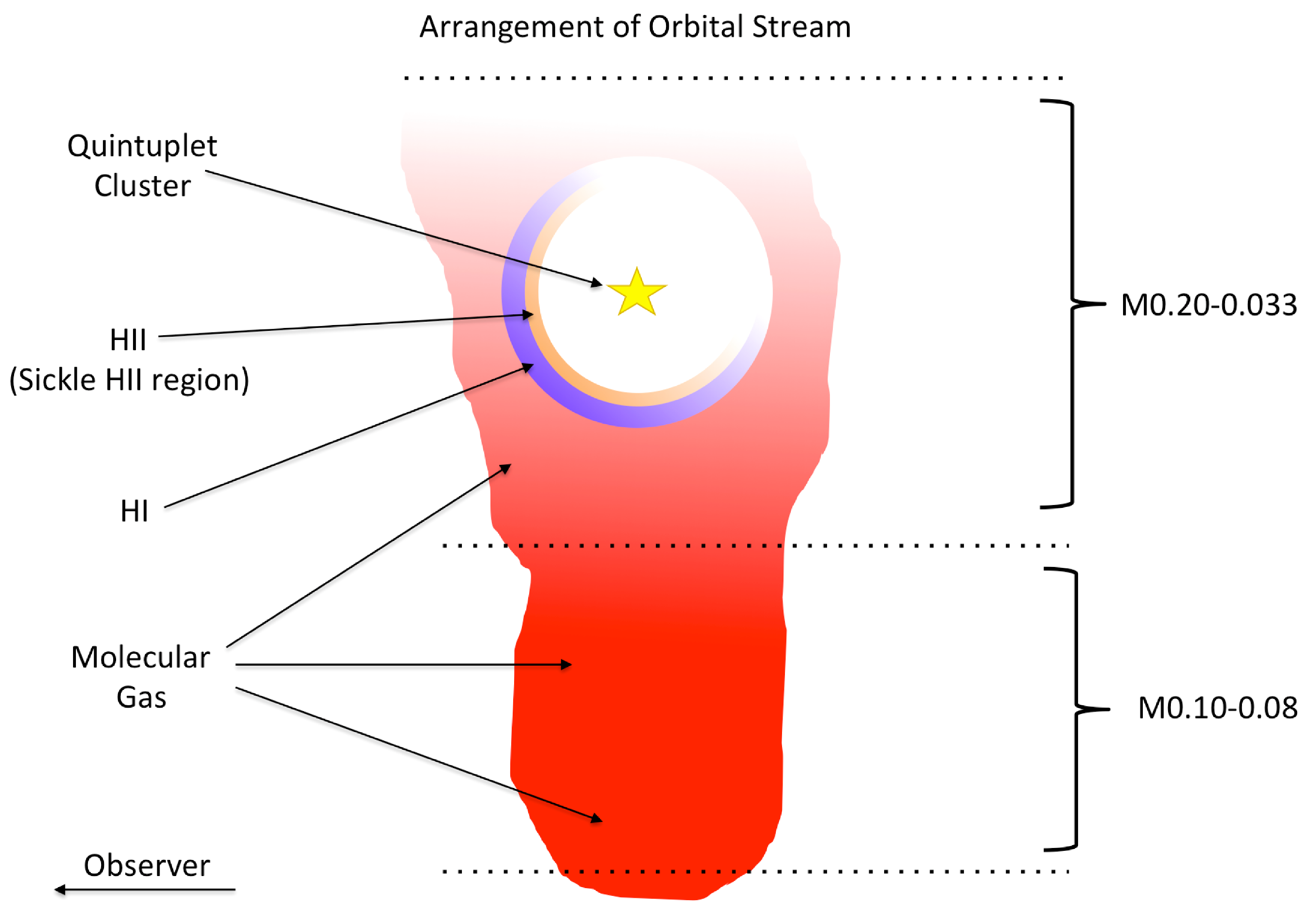}
\caption{Schematic depicting our suggested arrangement of molecular components in the orbital stream \citep[stream 1;][]{Kru15}, connecting \scloud~to \gcloud~as discussed in Section \ref{ion+hi}. The molecular gas is shown in red, the \hi~gas is in purple, and the ionized gas is in orange. The yellow star shows the location of the \qc.  }
\label{final}
\end{figure*}

Hot \xray~emission is present inside the \irbubble~\citep{ponti15}. \cite{ponti15} suggest that the hot gas inside this bubble is the combined result of winds from massive stars in the \qc~and multiple supernovae explosions. This theory is consistent with our momentum calculations for the \mshell, which indicate that multiple supernovae are required to produce the shell. 


\subsubsection{The Sickle \hii~Region: an Inner Ionized Layer}
\label{ion-dis}

As originally proposed by \cite{SG91} and \cite{lang97}, and supported by our observations here, \sick~appears to be the ionized edge of the \scloud~\mc~by the \qc. 
This orientation of the ionizing cluster, \hii~region, and \mc~is consistent with predicted models of expanding shells produced by star clusters \citep{weaver77, Dent09}.
Figure \ref{final} shows a possible arrangement of the interstellar and stellar components in the \rar~where \sick~represents an inner ionized layer between the cluster and the cloud. 
The bent morphology of \sick~and location inside the \mshell~(Figure \ref{sickle-shell-finder}), supports this suggestion. 
 Additionally, 21 $\mu$m dust emission \citep[MSX Band E;][]{Price01} is also detected \til1\arcmin~inside the eastern Galactic rim of the \mshell~(see Figure \ref{bubble}).  Further, this relatively bright dust emission closely follows \sick. Although no velocity information is obtained from these MSX observations, warm dust is known to be associated with \hii~regions \citep{Pal12}. 
 While the morphology of the thermal radio continuum and warm dust support the hypothesis of an inner ionized gas layer, our VLA data can reveal some additional position velocity information. We can examine the distribution of ionized gas in position-velocity space along the same slices (where we have data from \sick) as in Figure \ref{sickle-shell-pv}.

Given the proposed arrangement in Figure \ref{final}, we would expect to see that the velocities of the ionized gas would have values similar to the proposed position-velocity distribution for the expanding molecular shell (shown by the red dashed ellipses in Figure \ref{sickle-shell-pv}). 
The blue crosses in Figure \ref{sickle-shell-pv} show the approximate central velocity of the ionized gas (estimated to within 5 \kms) at several locations along the respective slice.
In nearly all cases, the values for the velocity of the ionized gas along slices I$-$III are close to the predictions for the velocities of expanding molecular shell. 
For example, the lowest ionized gas emission detected in slices I$-$III are located towards the predicted center of the shell, with values closer to the systemic velocity near the edge of the shell (Figure \ref{sickle-shell-pv}). This distribution supports the idea that the ionized gas is associated with the molecular shell.  

Slices I$-$III in Figure \ref{sickle-shell-pv} indicate that most of the ionized gas in \sick~have velocities lower than the systemic velocity of the \mshell~($v$ $\leq$ 53 \kms; Figures \ref{sickle-shell-pv}). 
These low velocities of the ionized gas are comparable with velocities detected in the low-velocity molecular component (\eg, see Figure \ref{sicklepv}), indicating the two components are physically connected. The similar velocities between most of the ionized gas and the low-velocity molecular component further suggests that the ionized gas is located on the nearside of the molecular shell. This orientation of components is consistent with the \hi~observations from \cite{lang10} (see Section \ref{swag shell}).
Additionally, the correlation between most of the ionized gas and \low~molecular component is consistent with the analysis in \cite{SG91} and \cite{lang97}.

While most of the detected ionized gas emission is associated with the low-velocity component, we also detect relatively bright ionized emission from a high velocity clump (\til75 \kms) that is located in \sickp~(see Figure \ref{ra mole}). Slice II in Figure \ref{sickle-shell-pv} further shows that the location of this clump in \pv~space correlates to molecular emission in the high-velocity component. The detection of ionized gas correlated with the high-velocity component supports our analysis that both molecular components of \scloud~are physically associated with the \qc. Thus, the distribution of the molecular and ionized gas, in \pv~space, further supports our suggestion that the ionized gas is an inner layer between the cluster and cloud, as illustrated in  Figure \ref{final}. 

Even though little ionized gas is detected in the \high~component (see Figure \ref{sickle-shell-pv}), the results of \cite{martin08} support the hypothesis that the \high~component is associated with the \qc. They observed low HNCO/$^{13}$CS abundance ratios in both velocity components, compared to other regions in the \gc, and they therefore argue that both velocity components have photodissociation regions (PDRs). These low abundance ratios are typically found in regions around massive star clusters as the HCNO molecule is destroyed in strong UV radiation fields. \cite{martin08} find a factor of 3 difference in the abundance ratios between the \loww~and \high~components, with the \high~component having a higher value. They attribute this difference to a slight offset in the physical distance of the two components to the ionizing source, concluding that the \low~component is physically closer to the \qc. These findings are consistent with our results of the ionized gas distribution, which show considerably less ionized emission in the high-velocity component than in the low-velocity component. 

The suggestion that \sick~is an inner ionized layer on the \mshell~is also supported by the data of \cite{langer17}. 
They observed \cii~towards the \gc~and argue that the \cii~emission within the inner 0$\fdg$25 of the GC is from highly ionized gas.
Additionally, Figure 7 in \cite{langer17} shows an elliptical ring in \pv~space at the \rar~(\eg, Figure \ref{orbitalmodel}, inset).\footnote{Note that our low-resolution \am~position-velocity slice (Figure \ref{SDfig}) is centered at a latitude of $b=-$0$\fdg$075, and covers a width of 0$\fdg$15. Therefore, the location of the \cii~slice in \cite{langer17} Figure 7, with a latitude of $b=0\fdg0$, is covered by the \am~slice. }
The velocity of the \cii~emission in this region range from 20 to 80 \kms~on both sides of the \pv~ellipse. This range of velocities detected in the ionized gas emission \citep{langer17} are consistent with the molecular gas velocities detected in this study.
This elliptical ring in the \cii~emission is located in \pv~space between streams 1 and 3 in the \cite{Kru15} orbital model and overlaps with molecular emission from both velocity components. 
As \cii~is also a known tracer of PDRs, the distribution of \cii~in \cite{langer17} is consistent with the results from \cite{martin08}, who argue that both velocity components of \scloud~contain evidence of PDRs.
Therefore, the consistency in the location and velocities of \cii~and the molecular gas indicates an inner PDR region (\cii) between the cluster and molecular gas, as depicted in Figure \ref{final}.

The presence of ionized gas, most likely on the inside of the shell, may help to explain the lack of molecular emission detected towards the north-eastern side of the molecular shell (Figure \ref{sickle-shell-finder}). 
The lack of molecular emission near the eastern side is also evident in single-dish observations by \cite{SG91} (see their Figure 2). However, as stated in Section \ref{shell-text}, there is molecular emission towards the north-eastern side of the shell at slightly lower and higher velocities (see the 40$-$50 \& 60$-$70 \kms~panels in Figure \ref{atca-panels}). 
Ionized gas emission is detected on the north-eastern Galactic side of \sick~(\ie, \sickp) at velocities around 50 \kms~(see Figure \ref{ra mole}, bottom right). The central velocities of the ionized gas in this region are comparable to the systemic velocities of \gcloud~and the \mshell. The spectral profiles of the ionized and molecular gas lines in Figure \ref{NH3spec}, \rega, also support this suggestion. The profile of the ionized gas emission in \sickp~shows an intermediate-velocity that bridges the velocity space between the two molecular components (see Section \ref{connection}). 

The potential age we derive for the \mshell~(1.3$\times$10$^{5}$ yr; Section \ref{shell-compare}) is also comparable to the timescales needed for pillar formation by ionizing radiation from a single O-type star \citep[2$-$4$\times$10$^5$ yrs;][]{freyer03, Mackey10,Stolte14}. 
In conclusion, the morphological, chemical, and kinematic features detected in this region support a model in which \sick~represents a foreground inner layer between the \scloud~expanding molecular shell~and the \qc.


\section{Conclusions}
\label{conclusion}

We present high-resolution (5\arcsec) radio observations made with the VLA of molecular (\am) and ionized gas (\rrl~radio recombination lines) in the Radio Arc region of the Galactic center. This paper focuses on the region around \scloud, \sick, and the \qc, but also addresses their physical relationship to other molecular clouds in the region (\eg, \gcloud). These high-resolution observations were compared with intermediate-resolution (\til20\arcsec; ATCA) and low-resolution (\til2\arcmin; Mopra) observations to yield the following conclusions: 

(1) The low and high-velocity components in M0.20-0.033 are physically connected in \pv~space toward the Galactic western side of the cloud, via an intermediate-velocity component (around 55 \kms). The connection between both velocity components is also observed on larger scales (\til2\arcmin; \til5 pc).   

(2) In the orbital model proposed by \cite{Kru15}, the kinematics of \scloud~place this complex on their stream 1. The physical connection in \pv~space between \scloud~and \gcloud~is consistent with this suggestion.

(3) We interpret the morphology and kinematic structure of \scloud~as an expanding shell. In this scenario, the \low~component in the ionized and molecular gas lies on the foreground side of the shell and the \high~component on the backside. 

(4) The \scloud~expanding shell has a systemic velocity~of 53 \kms, comparable to the central velocities detected in \gcloud, and a radius of $r$=135\arcsec~(5.2 pc). The expansion velocity of the shell is 40 \kms,~implying an upper limit on the age of \til1.3$\times$10$^5$ years.

(5)  The origin of the proposed expanding shell is located near the \qc, which is also on stream 1. 
The connection between the \qc~and the \mshell~is supported by our momentum calculations (Section \ref{ion+hi}). These show that \til30 supernovae, or a few hypernovae, are needed to power the shell. The expansion of the \scloud~shell~by multiple supernovae is analogous with other molecular shells in the \gc.
 
\section{Acknowledgements}

The research presented in this paper was supported by grants from the National Science Foundation (no. AST-0907934 and no. AST-15243000) and the NASA Iowa Space Grant Consortium (ISCG) 2015-2016 fellowship. Support for this work was also provided by the NSF through the Grote Reber Fellowship Program administered by Associated Universities, Inc./National Radio Astronomy Observatory. We would also like to thank Diederik Kruijssen, Miguel Requena-Torres, Allison Costa, and Sarah Sadavoy for their helpful discussion on this work. We would also like to thank the anonymous referee for reviewing this work. 

\software{CASA \citep{2011ascl.soft07013I}}


\appendix
\counterwithin{figure}{section}
\counterwithin{table}{section}
\section{Geometry of an Expanding Shell}
\label{app}

\subsection{Projected Radial Size of an Expanding Shell}
\label{los-geo}

\begin{figure}[hb]
\centering
\includegraphics[scale=0.6]{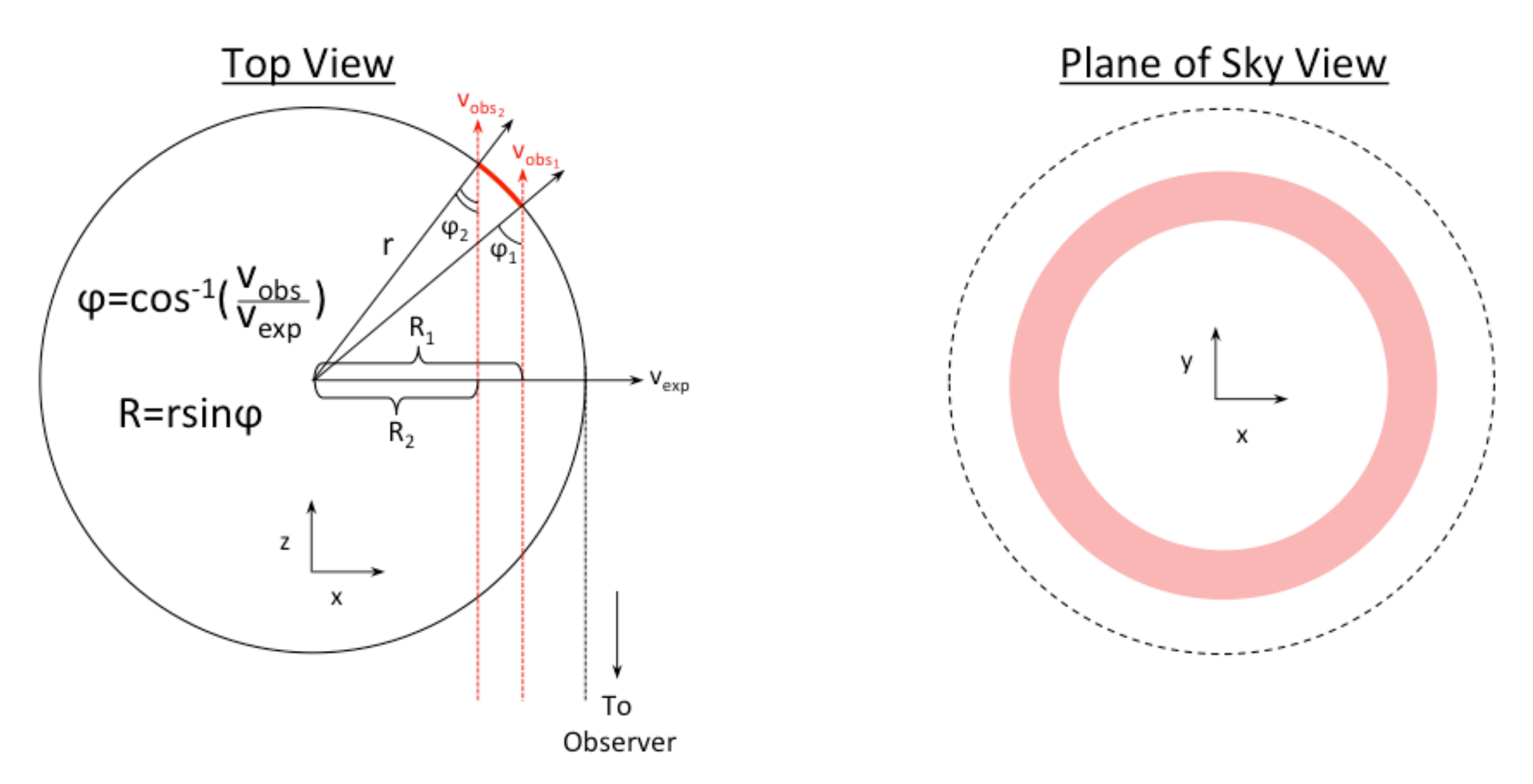}
\caption{(\leftit): Schedmatic illustrating the line-of-sight geometry of a 
uniformly expanding shell, where v$_{obs}$ is the observed doppler-shifted velocity, for an expansion velocity of v$_{exp}$, and R is the projected radius of the shell at v$_{obs}$. The two equations in \leftit~show the calculation of R using v$_{obs}$ and v$_{exp}$. (\rightit):~shows the plane of sky orientation of \leftit.} 
\label{radial-size}
\end{figure}

The apparent radial size of an expanding shell, projected in the sky, changes as a function of the observed Doppler-shifted velocity: v$_{obs}$. 
Figure \ref{radial-size} (left) shows the line of sight geometry of an expanding shell viewed at different Doppler-shifted velocities. In this schematic we are assuming that (1) we are in the reference frame of the expanding shell, (2) the shell is thin compared to the radius ($\Delta$r$<<$r), and (3) the shell is expanding isotropically in all directions. 
Under these assumptions, v$_{obs}$ changes as a function of $\varphi$, where $\varphi$ is the angle normal to the radial expansion vector: v$_{exp}$. 
The observed location of emission with v$_{obs}$ (R in Figure \ref{radial-size}, left), is also dependent on the sine of $\varphi$, for a fixed shell radius: r. For observed Doppler-shifted emission between v$_{obs1}$ and v$_{obs2}$, the projected radial size of the shell would range from R$_{1}$ to R$_{2}$ (Figure \ref{radial-size}, left). 

Figure \ref{radial-size} (right) shows the plane of sky orientation of the schematic depicted on the left. 
The black dashed circle represents the radial size of the shell at the systemic velocity (\ie, v$_{obs}$=0; see black dashed line in Figure \ref{radial-size}, left). The red shaded annulus shows the projected size of the two Doppler-shifted velocities from Figure \ref{radial-size}, left.
When viewed in the plane of the sky, gas between these velocities would form an annulus with an outer radius, R$_{1}$, and inner radius, R$_{2}$ (\eg, Figure \ref{radial-size}, right).

\subsection{An Expanding Shell in \PV~Space}
\label{pv-geo}

\begin{figure*}[hb]
\includegraphics[scale=0.64]{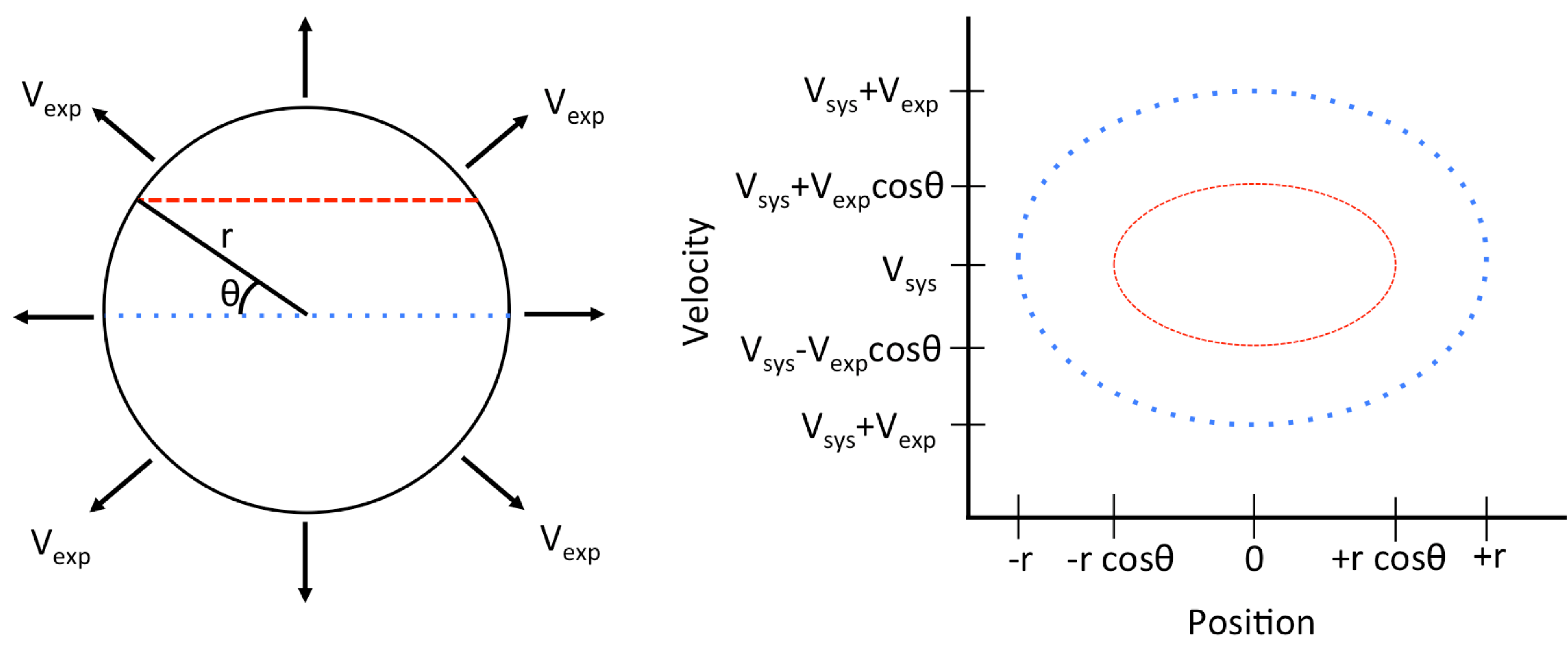}
\caption{Schematic depicting a translation of slices across an expanding shell into \pv~space. ({\it left}): illustration of the expanding shell, with a systemic velocity of \vsys, an expansion velocity of \vexp, and radius $r$.  ({\it right}): \pv~diagram of an expanding shell for the two different cuts across the shell on the {\it left}.}
\label{shell-model}
\end{figure*}

An expanding shell is known to produce an elliptical distribution in \pv~space. However, because the expanding shell is viewed in 2D space, the \los~expansion velocity detected is dependent on the $x$$-$$y$ coordinate of the shell in the plane of the sky. Figure \ref{shell-model} (left) illustrates an observed shell in the plane of the sky that has radius, $r$, expansion velocity, \vexp, and systemic velocity, \vsys. The red and blue dotted lines in Figure \ref{shell-model} (left) represent potential `slices' across the shell. Figure \ref{shell-model} (right) shows a \pv~schematic of these two example slices. The position axis (Figure \ref{shell-model}, right) corresponds to the length across the shell (Figure \ref{shell-model} left), or chord length, where 0 denotes the midpoint of the chord. The total length of any chord across the shell is equal to the major axis of the ellipse (\ie, 2r$\cdot$cos$\theta$). The blue slice (hereafter, slice 1) corresponds to a chord length equal to the diameter of the shell (\ie, $\theta$=0). The measured minimum and maximum velocities for slice 1 would reflect the expansion velocity (\vexp) as the vector is oriented parallel to our \los. The red slice (hereafter, slice 2) is located at some position angle, $\theta$, above (or below) a slice through the origin (\ie, slice 1).  At the midpoint of slice 2, the expansion velocity vector is also at an angle $\theta$ from our \los. Therefore the maximum Doppler shift of the gas we would detect along slice 2 is dependent on the cosine of the angle. For a uniformly expanding spherical shell, this schematic can be applied to any slice across the shell. 

\bibliographystyle{aasjournal}
\bibliography{Sickle.bib}

\end{document}